\documentclass[twocolumn,twoside]{IEEEtran}

\ifCLASSINFOpdf

\else

\fi

\ifCLASSOPTIONcompsoc
    \usepackage[caption=false, font=normalsize, labelfont=sf, textfont=sf]{subfig}
\else
    \usepackage[caption=false, font=normalsize]{subfig}
\fi
\usepackage{lipsum}%
\usepackage[dvipsnames]{xcolor}
\usepackage{algorithm,algorithmic}
\usepackage{balance}
\usepackage{multicol}   
\usepackage{cite}
\usepackage{gensymb}
\usepackage{multirow}
\usepackage{graphics}
\usepackage{epsfig}
\usepackage{graphicx}
\usepackage{epstopdf}
\usepackage{textcomp}
\usepackage{amsmath}
\usepackage{mathtools}
\usepackage{filecontents}
\usepackage{lipsum,color}
\usepackage{amssymb}
\usepackage{float}
\usepackage{colortbl} 
\usepackage{times} 
\usepackage{amsthm}  

\usepackage{amsfonts}

\usepackage{arydshln}

\theoremstyle{break}

\begin{document}
\title{Downlink Interference Analysis of UAV-based mmWave Fronthaul for Small Cell Networks}


\author{ Mohammad~Taghi~Dabiri,~Mazen~Hasna,~{\it Senior Member,~IEEE},~and~Walid~Saad,~{\it Fellow,~IEEE}
	\thanks{Mohammad Taghi Dabiri and Mazen Hasna are with the Department of Electrical Engineering, Qatar University, Doha, Qatar  (E-mail: m.dabiri@qu.edu.qa and hasna@qu.edu.qa).}
	\thanks{Walid Saad is with the Wireless$@$VT, Bradley Department of Electrical and Computer Engineering, Virginia Tech, Arlington, VA 22203 USA.}
	}
\maketitle
\vspace{-1cm}
\begin{abstract}
In this paper, an unmanned aerial vehicles (UAV)-based heterogeneous network is studied to solve the problem of transferring massive traffic of distributed small cells to the core network. First, a detailed three-dimensional (3D) model of the downlink channel is characterized by taking into account the real antenna pattern,  UAVs' vibrations, random distribution of small cell base stations (SBSs), and the position of UAVs in 3D space. Then, a rigorous analysis of interference is performed for two types pf interference: intra-cell interference and inter-cell interference. The interference analysis results are then used to derive an upper bound of outage probability on the considered system. Using numerical results show that the analytical and simulation results match one another. The results show that, in the presence of UAV's fluctuations, optimizing radiation pattern shape requires balancing an inherent tradeoff between increasing pattern gain to reduce the interference as well as to compensate large path loss at mmWave frequencies and decreasing it to alleviate the adverse effect of a UAV's vibrations. The analytical derivations enable the derivation of the optimal antenna pattern for any condition in a short time instead of using time-consuming extensive simulations.
\end{abstract}
\begin{IEEEkeywords}
Antenna pattern, backhaul/fronthaul links, interference, mmWave communication, small cell networks, unmanned aerial vehicles (UAVs).
\end{IEEEkeywords}
\IEEEpeerreviewmaketitle

\section{Introduction}
\IEEEPARstart{D}{riven} by the development of mobile Internet and smart phones, the need for data traffic increases by a factor of ten every five years and with the emerging Internet of Things (IoT) predicted to wirelessly connect trillions of users/devices in ultra-dense networks. \cite{saad2019vision,giordani2020toward,zeng2019accessing,mozaffari2019tutorial,khan2022digital}. 
%
%
%
%
%
%
With the dense deployment of small cell base stations (SBSs) which are vital for the ongoing network densification, high capacity backhaul/fronthaul links are required with a low latency of around 100 $\micro$s or less \cite{Ref20}. 
For the ongoing network densification, use of dense deployment of small cell base stations (SBSs) is vital which requires high capacity backhaul/fronthaul links \cite{Ref20}.
These demands can be traditionally fulfilled by coaxial cable or optical fiber in terrestrial networks. 
%
However, for massive deployment of small cells, such solutions will not be flexible and easy to deploy, and they won't be as cost-effective as compared to wireless backhaul/fronthaul links \cite{taori2015point}.
Wireless backhaul/fronthaul connectivity can be realized using microwave bands for non-line-of-sight (NLoS) cases or millimeter wave (mmWave) and free-space optical (FSO) links for line-of-sight (LoS) cases \cite{9492795}. Microwave links can cover a wide area but suffer from low data rates. FSO and mmWave-based backhaul/fronthaul links can meet the capacity requirements of next generation communication networks, and are lightweight and easy to install. 
However, mmWave/FSO links suffer from susceptibility to weather conditions \cite{alzenad2018fso} and require a LoS connection, which is a major challenge in urban regions due to the presence of small cells in hard-to-reach, near-street-level locations.
Recently, a promising alternative solution was presented in \cite{alzenad2018fso} that utilizes unmanned aerial vehicle (UAVs) as a wireless backhaul/fronthaul hub point between small cells and the core network. These UAV-hubs acting as networked flying platforms (NFPs) provide a possibility of wireless LoS fronthaul link.
One of the important challenges related to the considered NFP-based topology is the design of the fronthaul downlink using directional mmWave antennas which is the subject of this work. 
Although there are several works in this field \cite{rappaport2015wideband,ju2021millimeter,hemadeh2017millimeter,xu2021joint, gapeyenko2021line,          khawaja2017uav,khawaja2018temporal, 9426415,  
		tafintsev2020aerial,
		rupasinghe2019non,rupasinghe2019angle,yapici2021physical,
		gapeyenko2018flexible,
		kumar2020dynamic,dabiri2022enabling,dabiri2022study,dabiri2020analytical,dabiri20203d,dabiri20223d}, which will be reviewed in detail below, a comprehensive analysis of the considered downlink has not been provided in any of these studies by taking into account the effects of the real antenna pattern, the UAVs' fluctuations, the random distribution of SBSs, and the effects of inter-cell and intra-cell interference caused by side-lobes of real antenna patterns.
For example, in most of these studies \cite{gapeyenko2021line,khawaja2017uav,khawaja2018temporal, 9426415,tafintsev2020aerial,
	rupasinghe2019non,rupasinghe2019angle,  yapici2021physical,
	gapeyenko2018flexible,
	kumar2020dynamic}, the effect of the UAV's vibrations has been neglected, while we will show that by increasing the gain of the millimeter wave antenna, small fluctuations of less than one degree also affect the performance of the considered system. Moreover, in most of these studies, approximate patterns with fixed gain are used, which cannot obtain the exact results of the  performance.

\subsection{Literature Review and Motivation}
%
%
%
Although UAV-assisted channel modeling was investigated in recent studies \cite{khuwaja2018survey,nguyen2018novel,matolak2017air,sun2017air}, these works are mainly limited to sub-6 GHz frequency bands which cannot be directly employed to UAV-based mmWave communication systems. Meanwhile, most of the prior studies on mmWave channel modeling \cite{ rappaport2015wideband,ju2021millimeter,hemadeh2017millimeter} do not address the presence of UAVs, with the exception of a few recent works in \cite{xu2021joint, gapeyenko2021line,          khawaja2017uav,khawaja2018temporal, 9426415,  
	    tafintsev2020aerial,
	    rupasinghe2019non,rupasinghe2019angle,yapici2021physical,
	    gapeyenko2018flexible,
	kumar2020dynamic,dabiri2022enabling,dabiri2022study,dabiri2020analytical,dabiri20203d,dabiri20223d}.
For instance, a new LoS probability model for UAV-to-ground base station (BS) links is proposed in \cite{gapeyenko2021line} over realistic urban grid deployments in which the effects of building height distributions along with their densities and dimensions are considered.
In \cite{khawaja2017uav} and \cite{khawaja2018temporal}, the authors provided a new channel characterization for UAV-based mmWave links by using ray tracing simulations at two different frequency bands: 60 GHz and 28 GHz. 
In \cite{9426415}, a novel ground-to-UAV channel prediction method is provided by using ray tracing simulations based on the minimum Euclidean distance.
However, the results of these works \cite{gapeyenko2021line,khawaja2017uav,khawaja2018temporal, 9426415} are limited for omnidirectional mmWave antenna pattern.
%
%
High gain antennas can point the signals in a preferred direction in order to efficiently enhance the received signal power \cite{lu2020robust}.
Moreover, the narrow beamwidth of a directional antenna concentrates the signal power at the target point, which results in less interference and higher security in the physical layer.

Due to these advantages, the use of directional antennas for UAV-based mmWave communications was studied in \cite{tafintsev2020aerial,
	rupasinghe2019non,rupasinghe2019angle,  yapici2021physical,
	gapeyenko2018flexible,
	kumar2020dynamic,dabiri2022enabling,dabiri2022study,dabiri2020analytical,dabiri20203d,dabiri20223d}.
A dynamic algorithm is presented in \cite{tafintsev2020aerial} to get a better performance by considering dynamic backhauling and mobility of users.
A UAV-assisted configuration is introduced in \cite{rupasinghe2019non,rupasinghe2019angle,yapici2021physical} wherein a non-orthogonal multiple access technique was  employed to provide wireless connectivity for a large number of users. 
%
The use of aerial relay nodes was investigated in \cite{gapeyenko2018flexible} for both wireless access and backhaul mmWave links to allow dynamic routing.
A sectoring method is proposed in \cite{kumar2020dynamic} to guarantee ground user downlink coverage using a mmWave-enabled directional UAV link.
However, the results of the works in \cite{tafintsev2020aerial,
	rupasinghe2019non,rupasinghe2019angle,  yapici2021physical,
	gapeyenko2018flexible,
	kumar2020dynamic} assume that the UAVs are highly stable and not affected by vibrations. However, in practice, as the antenna gain increases, the sensitivity to the small UAV's vibration increases. 
%
This leads to an unreliable communication  due to antenna gain mismatch \cite{guan2019effects,zhong2019adaptive,pokorny2018concept}.

More recently, in \cite{dabiri2022study,dabiri2020analytical,dabiri20203d}, the authors studied the relationship between UAV's vibrations and mmWave antenna gain. The results of these work clearly show that the performance of a UAV-based link with directional mmWave antenna is highly dependent on the strength of UAV's vibrations. 
However, the results of  \cite{dabiri2022study} and \cite{dabiri2020analytical} are limited to a special case that does not account for the effects of interference caused by the side lobe gain of antenna array. 
More recently,  a UAV-assisted mmWave channel model was proposed in \cite{dabiri20223d} for a UAV-based network by taking into consideration real parameters such as UAV’s vibrations and interference caused by antenna side lobes. 
Generally, the dominant network traffic is for downloading high-quality videos and watching live sports matches or concerts. Therefore, the downlink capacity is often several times greater than that of the uplink and thus, its design is more important.
However, the results of \cite{dabiri20223d} are limited to the uplink and cannot be directly used for the design and analysis of a downlink.
Due to the asymmetric nature of UAV-based network, as we show in this study, the interference model of the uplink and downlink will be radically different. 
In addition, in \cite{dabiri20223d}, the questions regarding the effect of random distribution on SBSs have not been answered. 
%

\subsection{Contributions}
%
The main contribution of this paper is a holistic interference analysis of fronthaul downlink transmissions for a UAV-based network in the presence of the realistic channel parameters. To the best of the authors' knowledge, there exists no prior work on UAV-based mmWave downlink communications that jointly takes into account the actual antenna pattern, the UAVs' vibrations, and the random effect of the SBSs. In summary, the paper makes the following key contributions:

\begin{itemize}
	\item First, we characterize a detailed 3D modeling of the downlink channel, by taking into account the real antenna pattern,  UAVs' vibrations, random distribution of SBSs, and the position of NFPs in 3D space. Using this characterization, we derive the end-to-end signal-to-noise plus interference ratio (SINR) by considering all the mentioned parameters.
	%
	\item We then rigorously analyze the interference in the system. Interference is divided into two important categories: intra-cell interference and inter-cell interference. We model both types of interference by considering all channel parameters. Then, using simulations, the effect of the channel parameters on the interference is studied and important results are extracted regarding the reduction of the interference effect.
	\item 
	Subsequently, we use the obtained results about interference in order to examine the system performance in terms of outage probability. To this end, first, the effect of the random distribution of SBSs on the outage probability is studied.
	We will show that the important parameter affecting the interference is the spatial angle between the mmWave directional antennas, which is a function of the random distribution of the SBSs. Therefore, we derive the probability density function (PDF)  and cumulative distribution function (CDF) of spatial angle between the mmWave directional antennas. Using those results, we derive an upper bound on the average outage probability of the considered system by taking into account all the channel parameters.
	\item 
	Finally, using extensive simulation and numerical results, we evaluate the performance of the considered system in terms of outage probability. Our simulation results show that the derived analytical expression is close to the simulated system with acceptable accuracy. 
	We also show that in the presence of UAV's fluctuations, it is very necessary to choose the optimal antenna pattern to manage the interference and at the same time to reduce the effect of UAV's fluctuations. Our analytical derivations allow us to obtain the optimal antenna pattern for any condition in a short time instead of using time-consuming extensive simulations.
\end{itemize}

The rest of this paper is organized as follows. We characterize the considered downlink channel in Section II.
Interference modeling and analysis are done in Section III, and outage probability analysis along with related simulations are provided in Section IV.
Finally, conclusions are drawn in Section V.

\begin{figure*}
	\centering
	\subfloat[] {\includegraphics[width=2.7 in]{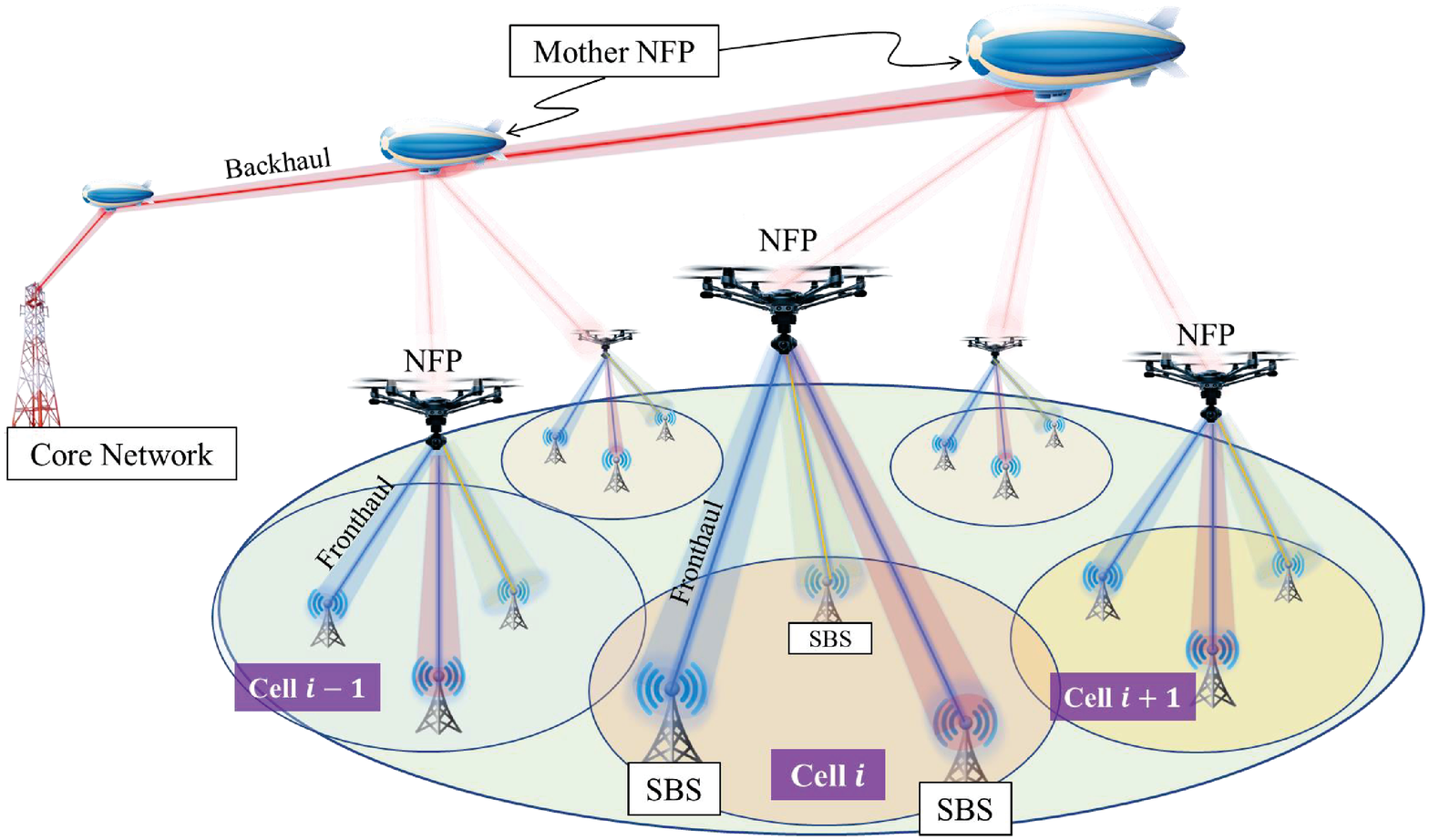}
		\label{xn1}
	}
	\hfill
	\subfloat[] {\includegraphics[width=2.1 in]{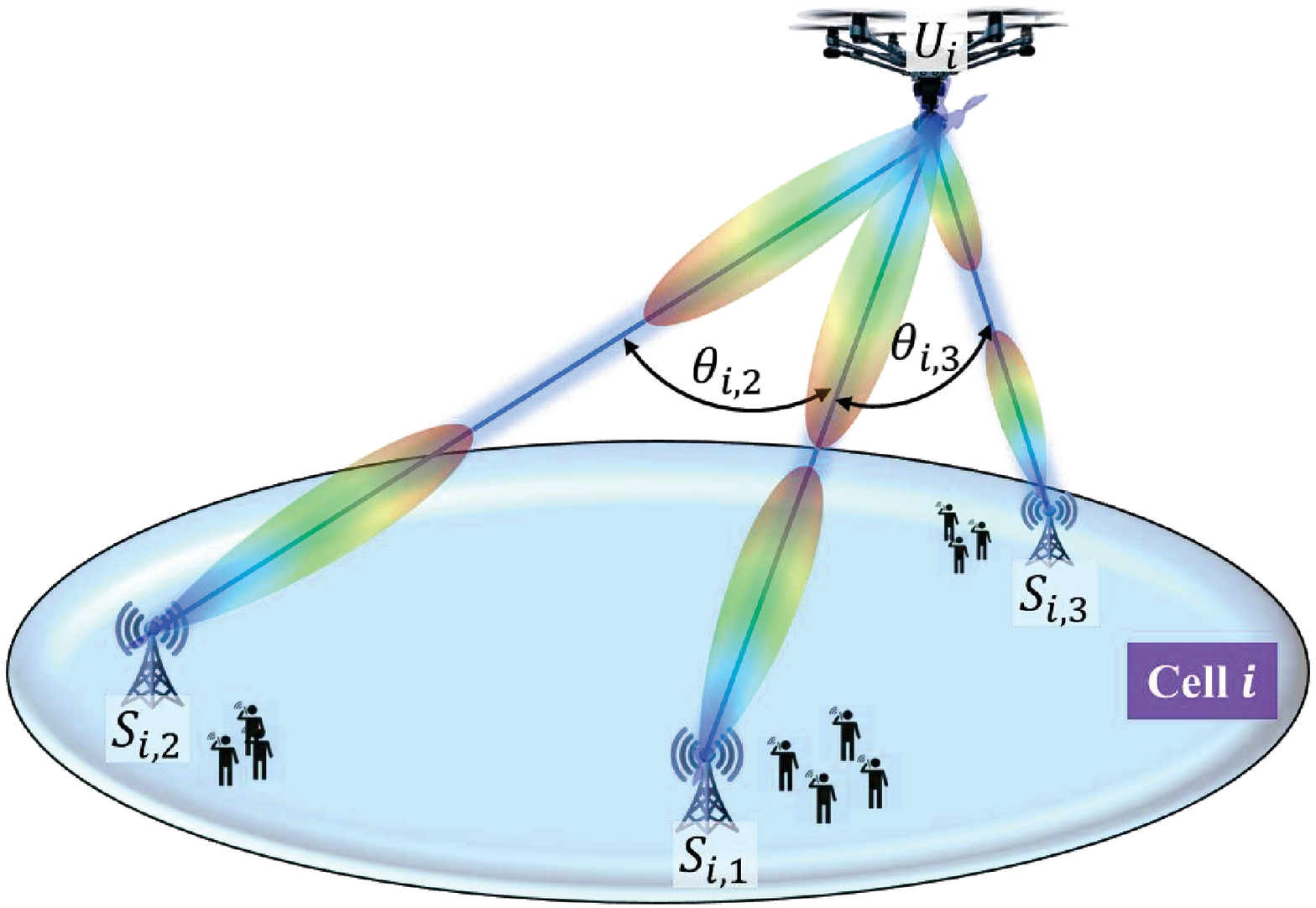}
		\label{xn2}
	}
	\hfill
	\subfloat[] {\includegraphics[width=2.1 in]{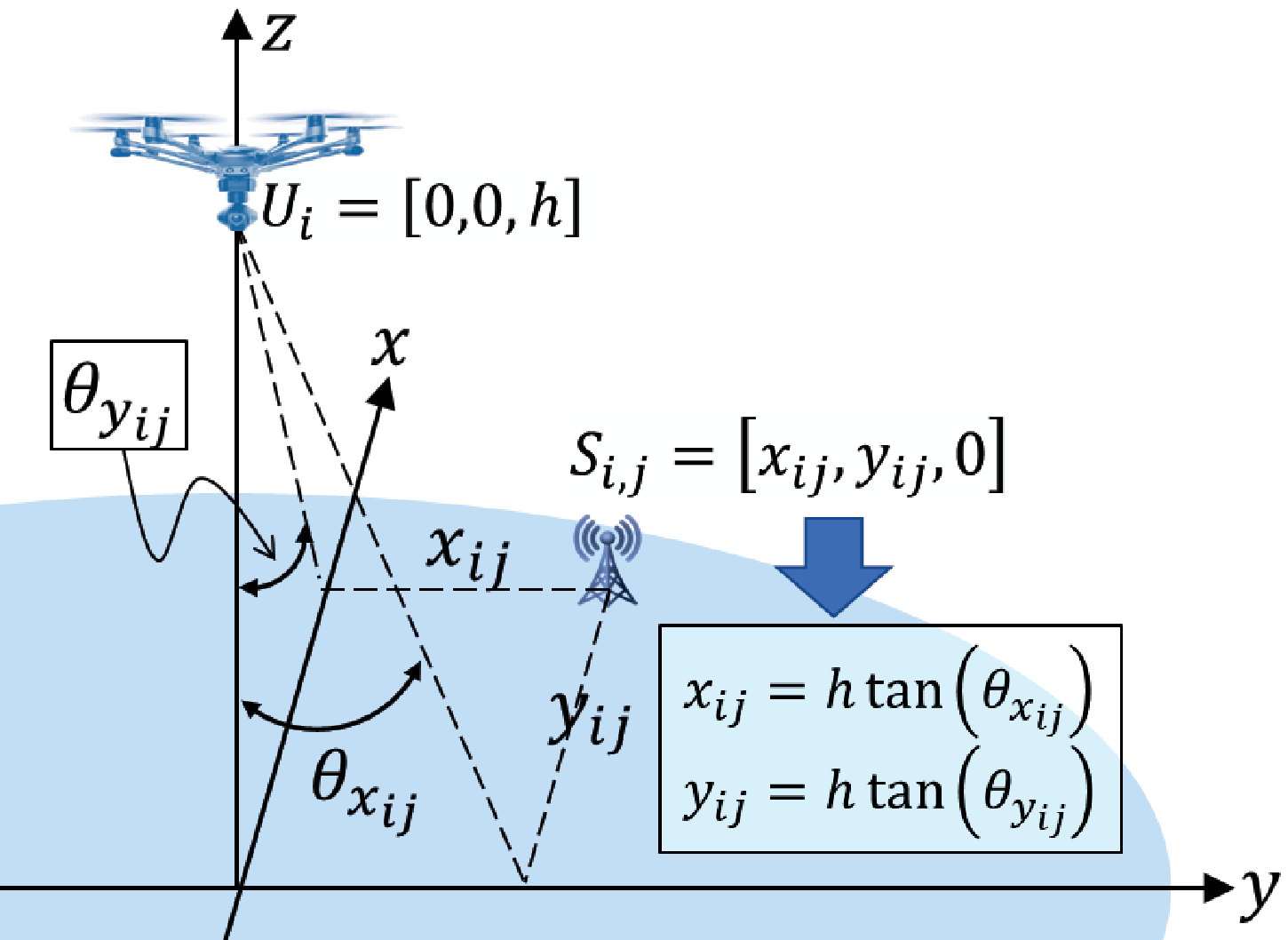}
		\label{xn3}
	}
	\caption{(a) An illustration of a UAV-assisted HetNet as an alternative solution for backhaul/fronthaul links
		which uses UAVs and higher directional frequencies (such as mmWave, THz, and optical frequencies) to transfer traffic from the distributed SBSs to the core network.
		In this topology, each UAV is able to establish fronthaul links for several SBSs.
		(b) A 3D illustration of cell $i$  for which several SBSs (denoted by $\mathcal{S}_{i,j}$ for $j\in\{1,...,J\}$) are connected to $U_i$. The spatial angle between the downlink transmissions of $\mathcal{S}_{ij}$ and $\mathcal{S}_{i1}$ is denoted by $\theta_{ij}$. Here, higher values for $\theta_{ij}$ or lower values for antenna beamwidth lead to a lower interference between $\mathcal{S}_{ij}$ and $\mathcal{S}_{i1}$.
		The position of each $\mathcal{S}_{ij}$ relative to the $U_i$ is determined using angles $\theta_{x_{ij}}$ and $\theta_{y_{ij}}$.}
	\label{xn4}
\end{figure*}

\begin{table}
	\caption{The list of main notations.} 
	\centering 
	\begin{tabular}{l l} 
		\hline\hline \\[-1.2ex]
		{\bf Parameter} & {\bf Description}  \\ [.5ex] 
		\hline\hline \\[-1.2ex]
		Subscript $i$ & Denote the cell $i$\\
		Subscript $j$ & Denote the SBS $j$ in each cell \\
		Subscript $q$ & $\in\{s,u\}$ determines the SBS and UAV nodes \\
		Subscript $w$ & $\in\{x,y\}$ determines $x$ and $y$ axes \\
		%
		%
		$\sigma_\theta$ & Characterize the strength of instability of UAVs\\
		$J$ & Number of SBSs in each cell \\
		$I$ & Number of UAVs \\
		$\alpha_c$ & Defined in Remark 5 \\
		$U_i$ & Denote $i$th NFP where $i\in\{1,...,N_D\}$ \\
		$\mathcal{S}_{ij}$& Denote $j$th SBS in the $i$th cell\\
		$T_{ij}$ & Denote the link between $U_i$ and $\mathcal{S}_{ij}$\\
		%
		$A_{\mathcal{S}_{ij}}$ & Denote the array antenna of $\mathcal{S}_{ij}$\\
		$A_{u_{i}}$ & Denote the array antenna of $U_i$\\
		$P_{t}$ & Transmitted power \\
		$L_{ij}$ & Link length of $T_{ij}$ link \\
		$h_{L_{ij}}$ & Path loss of a $T_{ij}$ link \\
		$N_{u}$ & Denote  square array antenna of $U_i$ with \\[-.7ex]
		&$N_{u}\times N_{u}$ elements\\
		$N_{s}$ & Denote  square array antenna of $\mathcal{S}_{ij}$ with \\[-.7ex]
		&$N_{s}\times N_{s}$ elements\\ 
		$h_{i}$ & The height of $i$th NFP\\
		$[x; y; z]$ & Cartesian coordinate system that the position of each $\mathcal{S}_{ij}$  \\[-.7ex]
		& and $U_1$ are characterized as $[x_{ij},y_{ij},0]$ and $[0,0,h_1]$ \\ [.5ex]
		$\lambda$ and $f_c$ &  Wavelength and carrier frequency \\
		$\alpha$ nad $\beta$ & constants whose values depend on  the\\[-.7ex]
		&  propagation environment\\[.5ex]
		$\sigma^2_{N}$ &  The thermal noise power \\ 
		%
		%
		%
		\hline
		$\Theta_i$ &  $=[\Theta_{x_i},\Theta_{y_i}]$ that indicate the orientation fluctuations \\[-.5ex]
		& of $U_i$ in $x-z$, and $y-z$ planes. See Fig. \ref{nv1} \\
		%
		%
		%
		$\theta_{w_{ij}}$  & Defined in \eqref{p5} and Fig. \ref{xn3} where $w\in\{x,y\}$ \\ 
		$\theta_{ij}$  & It is calculated from $\theta_{x_{ij}}$ and $\theta_{y_{ij}}$, similar to \eqref{f_2} \\ 
		$\theta_{wd_{1j}}$ & $=(\theta_{w_{11}}-\theta_{w_{1j}})$. See Fig. \ref{nv1}. \\
		$\theta_{d_{1j}}$ & It is calculated from $\theta_{xd_{1j}}$ and $\theta_{yd_{1j}}$, similar to \eqref{f_2} \\
		$\theta_{\textrm{elev},i}$ & Elevation angle of $\mathcal{S}_{11}$ compared to $U_i$\\
		%
		%
		%
		\hline 
		$f_x(x)$ & The PDF of RV $x$ \\
		$F_x(x)$ & The CDF of RV $x$ \\
		%
		%
		%
		%
		\hline \hline              
	\end{tabular}
	\label{I1} 
\end{table}

\section{The System Model}
%
As shown in Fig. \ref{xn1}, we consider a UAV-assisted HetNet  that is offered as an affordable and easy to deploy option in \cite{alzenad2018fso} to solve the concern related to transferring traffic of the distributed SBSs to the core network. More precisely, as an alternative for backhaul/fronthaul links, the UAV-based mmWave or FSO links are suggested in \cite{alzenad2018fso} to transfer the SBSs traffic, particularly in ultra-dense urban areas. 
However, under cloudy, foggy, and raining conditions, a FSO communication link experiences high attenuation and thus, it becomes difficult to guarantee the required backhaul/fronthaul quality-of-service (QoS) \cite{alzenad2018fso}. Unlike FSO links, mmWave links are more robust to foggy weather. Therefore, we use mmWave for backhaul/fronthaul links of the considered system. 
%
As seen in Fig. \ref{xn1}, the considered HetNet consists of several different wireless backhaul/fronthaul links: i) SBS-to-NFP (S2N) links, ii) inter NFP links, and iii) inter mother NFP links. 
Our focus will be on the modeling and design of the S2N links. For an S2N fronthaul link, the uplink consists of a transmission from a ground remote radio head (RRH)  to an NFP, and the downlink transmission will be essentially a link from an NFP to an RRH.
%
Here, we focus on the downlink.

\subsection{The 3D Antenna Pattern}
%
In order to reduce the effect of interference and also to reduce the negative effect of channel attenuation at higher frequencies, the use of high gain antenna is essential, particularly for ultra high data rate backhaul/fronthaul links. Advances in the fabrication of antenna array technology at higher frequency allow the creation of large antenna arrays with high gain in a cost effective and compact form \cite{ghattas2020compact,asaadi2018high}. 
In addition, as we will show later, employing a directional mmWave antenna pattern allows us to reuse frequency bands thus improving the spectral efficiency of the considered system. 

We assume that the antennas installed on the NFPs have the same pattern and the antennas installed on the SBSs have the same pattern as well.
We consider a uniform square array antennas for both NFPs and SBSs, comprising $N_q\times N_q$ antenna elements with the same spacing $d_a$ between elements, where the subscript $q=s$ determines the SBS node and the subscript $q=u$ determines UAV node. 
The array radiation gain is mainly formulated in the direction of $\theta$ and $\phi$. 
By taking into account the effect of all elements, the array radiation gain will be:
\begin{align}
	\label{p_1}
	G_q(N_q,\theta,\phi)  = G_0(N_q) \,
	\underbrace{G_e(\theta,\phi) \,  G_a(N_q,\theta,\phi)}_{G'_q(N_q,\theta,\phi)},
\end{align}
where $G_a$ is an array factor, $G_e$ is single element radiation pattern and $G_0$ is defined in \eqref{cv}. 
From 3GPP single element radiation pattern, $G_{e,\textrm{3dB}}=10\times\log_{10}(G_e)$ of each single antenna element will be given by \cite{niu2015survey}
\begin{align}
	\left\{ \!\!\!\!\! \! \!
	\begin{array}{rl}
		&G_{e\textrm{3dB}} = G_{\textrm{max}} - \min\left\{-(G_{e\textrm{3dB,1}}+G_{e\textrm{3dB,2}}),F_m 
		\right\},  \\
		&G_{e\textrm{3dB,1}} =  -\min \left\{ - 12\left(\frac{\theta_e-90}{\theta_{e\textrm{3dB}}}\right)^2,
		G_{\textrm{SL}}\right\},    \\
		&G_{e\textrm{3dB,2}} = -\min \left\{ - 12\left(\frac{\theta_{x}}{\phi_{e\textrm{3dB}}}\right)^2,
		F_m\right\}, \\
		&\theta_e             = \tan^{-1}\left( \frac{\sqrt{1+\sin^2(\theta_{x})}}
		{\sin(\theta_{y'})} \right),
	\end{array} \right. \nonumber
\end{align}
where $\theta_{e\textrm{3dB}}=65^{\circ}$ and $\phi_{e\textrm{3dB}}=65^{\circ}$ are the vertical and horizontal 3D beamwidths, respectively, $G_{\textrm{max}}$ is the maximum directional gain of the antenna element, $F_m=30$ dB is the front-back ratio, and $G_{\textrm{SL}}=30$ dB is the side-lobe level limit. 
If the amplitude excitation of the entire array is uniform, then the array factor $G_a(N_q,\theta,\phi)$ for a square array of $N_q\times N_q$ elements can be obtained as \cite[eqs. (6.89) and (6.91)]{balanis2016antenna}:
\begin{align}
	\label{f_1}
	G_a(N_q,\theta, \phi) &= 
	\left( \frac{\sin\left(\frac{N_q (k d_a \sin(\theta)\cos(\phi)+\beta_{x})}{2}\right)} 
	{N_q\sin\left(\frac{k d_a \sin(\theta)\cos(\phi)+\beta_{x}}{2}\right)}
	\right. \nonumber \\
	&\times \left. \frac{\sin\left(\frac{N_q (k d_a \sin(\theta)\sin(\phi)+\beta_{y})}{2}\right)} 
	{N_q\sin\left(\frac{k d_a \sin(\theta)\sin(\phi)+\beta_{y}}{2}\right)}\right)^2,
\end{align}
where $d_a=\frac{\lambda}{2}$ and $\beta_{w}$ are the spacing and progressive phase shift between the elements, respectively. $k=\frac{2\pi}{\lambda}$ is the wave number, $\lambda=\frac{c}{f_c}$ is the wavelength, $f_c$ is the carrier frequency, and $c$ is the speed of light. 
Also, in order to guarantee that the total radiated power of antennas with different $N$ are the same, the coefficient $G_0$ is defined as
\begin{align}
	\label{cv}
	G_0(N_q)=\frac{4\pi}{\int_0^{\pi}\int_0^{2\pi} G'_q(\theta,\phi) \sin(\theta) d\theta d\phi}.
\end{align}
Based on \eqref{f_1}, the maximum value of the antenna gain is equal to $G_0(N_q)$, which is obtained when $\theta=0$.
%

\subsection{Topology Description}
Let $U_i$ represent the $i$th NFP, and let $\mathcal{S}_{i,j}$ be the set of SBSs connected to $U_i$. Also, $A_{\mathcal{S}_{ij}}$ indicates the array antenna of $\mathcal{S}_{i,j}$, which is directed towards the $U_i$. 
As shown in Fig. \ref{xn2}, the set of $\mathcal{S}_{i,j}$ connected to $U_i$ forms the cell $i$.
We assume that the SBSs are distributed in the $x-y$ plane and the $U_i$ are flying at height $h_i$.
Therefore, the position of each $\mathcal{S}_{ij}$ is characterized as $[x_{ij},y_{ij},0]$ in a Cartesian coordinate system where $[x,y,z]\in\mathbb{R}^{1\times3}$.
As shown in Fig. \ref{xn3}, the position of each $\mathcal{S}_{ij}$ relative to the $U_i$ is uniquely determined using angles $\theta_{x_{ij}}$ and $\theta_{y_{ij}}$ as follows: 
\begin{align}
	\label{p5}
	x_{ij} = h_i \tan(\theta_{x_{ij}}),~~~y_{ij} = h_i \tan(\theta_{y_{ij}}).
\end{align}
Each $U_i$ is equipped with $J$ directional mmWave array antennas, which are denoted by $A_{u_{ij}}$. 
The direction of $A_{u_{ij}}$ can be electrically or using a simple stepper motor, aligned towards $A_{\mathcal{S}_{ij}}$ \cite{heath2016overview}.
Therefore, according to \eqref{p5}, UAV $U_i$ sets the direction of each of the antennas on the angle $\theta_{ij}=[\theta_{x_{ij}},\theta_{y_{ij}}]$.
We define $T_{ij}$ as the communication link between $A_{\mathcal{S}_{ij}}$ and $A_{u_{ij}}$. 

In order to analyze the considered system, we need a comprehensive channel model in which the effect of the spatial distribution of SBSs, the spatial distribution of NFPs, antenna patterns, and UAVs' instabilities are considered.
For notational simplicity, we obtain the SINR for link $T_{11}$ in general, which can be easily used to represent the rest of the links.
Note that we have two types of interference, one is intra-cell interference due to $T_{1j}$ links and the other is inter-cell interference due to $T_{ij}$ links where $i\neq1$. Therefore, the SINR at the receiver (Rx) of $\mathcal{S}_{11}$ will be:
\begin{align}
	\label{p1}
	\gamma_{11} = \frac{P_{r_{11}}}{ 
		\underbrace{\sum_{j=2}^J\mathbb{I}_{1j}}_\text{intra-cell} + 
		\underbrace{\sum_{i=2}^{I} \sum_{j=1}^J \mathbb{I}_{ij}}_\text{inter-cell interference} + \sigma^2_{N}}
\end{align}
where $\sigma^2_{N}$ is the thermal noise power. $P_{r_{11}}$ is the received power transmitted by $A_{\mathcal{S}_{11}}$ given by:
\begin{align}
	\label{p2}
	P_{r_{11}} = P_t |h_{L_{11}}|^2 
	\underbrace{G_{u_{11}}(N_{u_{11}},\Theta_{u_{11}},\phi_{u_{11}})}_\text{Pattern Gain of $A_{u_{ij}}$} 
	\underbrace{G_{\mathcal{S}_{11}}(N_{\mathcal{S}_{11}},\Theta_{\mathcal{S}_{11}},\phi_{\mathcal{S}_{11}})}_\text{Pattern Gain of $A_{\mathcal{S}_{11}}$}.
\end{align}
In \eqref{p2}, $P_t$ is the transmit power, $h_{L_{ij}}=h_{Lf_{ij}}h_{Lm_{ij}}$ is the channel path loss, $h_{Lf_{ij}}=\left(\frac{\lambda}{4\pi L_{ij}}\right)^2$ is the free-space path loss, $h_{Lm_{ij}}= e^{-\frac{\mathcal{K}(f)}{2}L_{ij}}$ represents the molecular absorption loss where $\mathcal{K}(f)$ is the frequency dependent absorption coefficient, and $L_{ij}$ is the link length of $T_{ij}$.
Also, random variables (RVs) $\Theta_{u_{ij}}$ and $\Theta_{\mathcal{S}_{ij}}$ capture the orientation fluctuations of $A_{u_{ij}}$ and $A_{\mathcal{S}_{ij}}$, and they are modeled in the next section. For simplicity, the simplified symbol $\Theta_i=[\Theta_{x_i},\Theta_{y_i}]$ is used to refer to RV $\Theta_{u_{ij}}$.
\subsection{Modeling UAV Vibrations}
In an ideal scenario, the directional antennas are perfectly aligned and the parameters $\Theta_{q_{ij}}$ in \eqref{p2} tend to zero and the received signal power tends to the maximum value as
\begin{align}
	\label{p3}
	P_{r_{11}} \simeq P_t |h_{L_{11}}|^2 G_0(N_u) G_0(N_s).
\end{align}
Based on the ideal results in \eqref{p3}, a higher value of $N_q$, leads to a higher value for the received power. Also, in Section III, we will show that a larger $N_q$ can lead to a lower interference, and hence, the SINR always improves by increasing $N_q$.
However, in real situations, the SINR does not necessarily improve by increasing $N_q$. 
In practical situations, an error in mechanical control system, mechanical noise, position estimation errors, air pressure, and wind speed can affect the UAV's angular and position stability.
In practice, the instantaneous orientation of a UAV can randomly deviate from its means denoted by $\Theta_q$. This, in turn, leads to deviations in the angle of departure (AoD) of the transmitter (Tx)'s antenna pattern.
Therefore, as $N_q$ increases, the system becomes more sensitive to the UAV's vibrations. Therefore, on the one hand, we must increase $N_q$ to reduce interference as well as to increase the desired received power. On the other hand, we must be careful not to choose a very large value for $N_q$ such that even with small fluctuations in antenna direction, the probability of missing the main lobe and being on the side-lobes increases.

The orientation fluctuations of antennas mounted on $U_i$ are captured by $\Theta_i=[\Theta_{x_i},\Theta_{y_i}]$ where $\Theta_{x_i}$ and $\Theta_{y_i}$ are the instantaneous fluctuations of $U_i$ in the $x-z$ and $y-z$ Cartesian coordinates, respectively.
Therefore, the overall instantaneous deviation angle of $A_{u_{ij}}$ is obtained as:
\begin{align}
	\label{f_2}
	\Theta_i  &= \tan^{-1}\left(\sqrt{\tan^2(\Theta_{x_i})+\tan^2(\Theta_{y_i})}\right), \nonumber \\
	\phi_i    &=\tan^{-1}\left(\frac{\tan(\Theta_{y_i})}{\tan(\Theta_{x_i})}\right).
\end{align}
Based on the central limit theorem, the UAV's orientation fluctuations are considered to be Gaussian distributed \cite{dabiri2018channel,kaadan2014multielement,dabiri2019tractable}. 
Therefore, we have 
$\Theta_{x_i}\sim \mathcal{N}(0,\sigma^2_{\theta_x})$, and $\Theta_{y_i}\sim \mathcal{N}(0,\sigma^2_{\theta_y})$.
For the ground SBSs, the estimation error of the exact position of the flying NFPs and the insufficient speed to track NFPs will cause an angular error. Unlike the flying NFPs, we assume that the ground SBSs do not face weight and power consumption limitations and, hence, they can align their antennas with the considered NFP with a negligible angle error. Therefore, \eqref{p2} can be rewritten as follows:
\begin{align}
	\label{p6}
	P_{r_{11}} = P_t |h_{L_{11}}|^2 G_0(N_s)
	G_{u}(N_{u},\Theta_1,\phi_1).
\end{align}

\subsection{Probability of LoS}
In addition to the high propagation loss, mmWave communication systems are very sensitive to blockages \cite{bai2014coverage}. 
%
Therefore, the probability of LoS is an important factor and can be described as a function of the elevation angle and environment as follows \cite{al2014modeling,al2014optimal}:
\begin{align}
	\label{op11}
	P_\textrm{LoS}(\theta_\text{elev}) = \frac{1}
	{1+\alpha \exp\left(-\beta(\frac{180}{\pi}\theta_\text{elev}-\alpha)\right)}	
\end{align}
where $\alpha$ and $\beta$ are constants whose values depend on the propagation environment, e.g., rural, urban, or dense urban, and $\theta_\textrm{elev}$ is the elevation angle.

\begin{figure}
	\begin{center}
		\includegraphics[width=3.4 in]{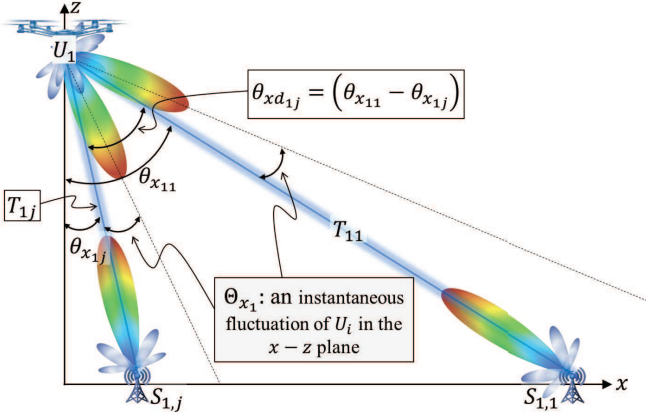}
		\caption{Graphical illustration of intra-cell interference in 2D $x-z$ plane caused by antenna pattern of $A_{u_{1j}}$.
			The instantaneous orientation of $U_1$ randomly deviates from its means $\Theta_{x_1}$ and $\Theta_{y_1}$ in the $x-z$ and $y-z$ planes.
		}
		\label{nv1}
	\end{center}
\end{figure}
%

\begin{figure}
	\centering
	\subfloat[] {\includegraphics[width=3.3 in]{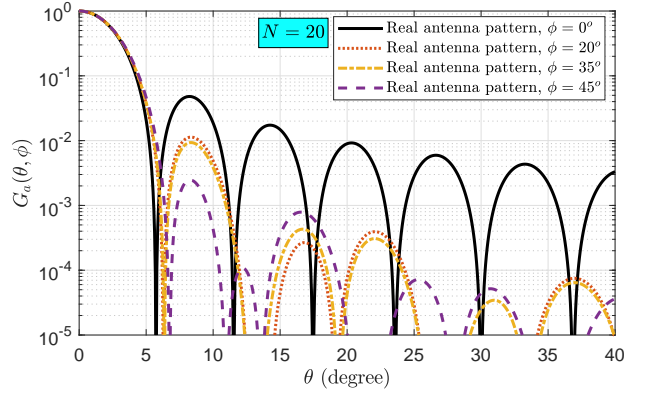}
		\label{nv4}
	}
	\hfill
	\subfloat[] {\includegraphics[width=3.3 in]{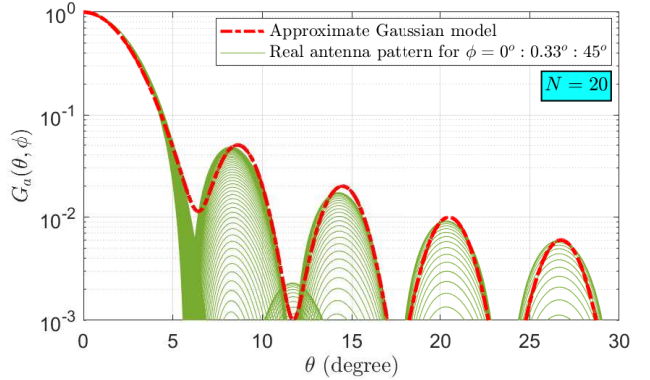}
		\label{nv5}
	}
	\caption{(a) Real antenna pattern of a $20\times20$ standard array antenna versus $\theta$ for different values of $\phi$.
		Comparing the accuracy of the model obtained from \eqref{c5} with the real antenna pattern model for different values of $\phi=0:0.33:45$.}
	\label{nv6}
\end{figure}

\section{Interference Analysis}
In this section, we first model interference and then investigate the effect of interference under different conditions.

\subsection{Interference Modeling}
For the considered system, the interference is divided into two categories, intra-cell and inter-cell interference, as modeled next.

\subsubsection{Intra-cell Interference}
For notation simplicity, here, we provide the general channel model for $T_{11}$ link which connects $U_1$ to $\mathcal{S}_{11}$ in presence of interference from other $\mathcal{S}_{ij}$ and $U_i$. The obtained model can be easily used for other $T_{ij}$ links.
For link $T_{11}$, the intra-cell interference is caused by the interference from other links $T_{1j}$ for $j\in\{2,...,J\}$. Therefore, as shown in Fig. \ref{nv1}, the intera-cell interference will be:
\begin{align}
	\label{n1}
	\mathbb{I}_\text{intra} = \sum_{j=2}^J\mathbb{I}_{1j},
\end{align}
where $\mathbb{I}_{1j}$ is the interference caused by $T_{1j}$ link. As graphically illustrated in Fig. \ref{nv1}, $\mathbb{I}_{1j}$ is obtained in \eqref{sd1} where  we have
$\theta_{xd_{1j}}=(\theta_{x_{11}}-\theta_{x_{1j}})$, and
$\theta_{yd_{1j}}=(\theta_{y_{11}}-\theta_{y_{1j}})$. 
\begin{figure*}[!t]
	\normalsize
	\begin{align}
		\label{sd1}
		\mathbb{I}_{1j} =&~ P_t |h_{L_{11}}|^2 G_0(N_s) G_0(N_u)    G_e\left(\tan^{-1}\left(\sqrt{\tan^2(\theta_{xd_{1j}}-\Theta_{x_1})+\tan^2(\theta_{yd_{1j}}-\Theta_{y_1})}\right)\right) \nonumber \\
		&\times\left( \frac{\sin\left(\frac{N_q (k d_a \sin\left(\tan^{-1}\left(\sqrt{\tan^2(\theta_{xd_{1j}}-\Theta_{x_1})+\tan^2(\theta_{yd_{1j}}-\Theta_{y_1})}\right)\right)
				\cos(\phi_{u_{1j}}))}{2}\right)} 
		{N_q\sin\left(\frac{k d_a \sin\left(\tan^{-1}\left(\sqrt{\tan^2(\theta_{xd_{1j}}-\Theta_{x_1})+\tan^2(\theta_{yd_{1j}}-\Theta_{y_1})}\right)\right)
				\cos(\phi_{u_{1j}})}{2}\right)}
		\right. \nonumber \\
		&\times \left. \frac{\sin\left(\frac{N_q (k d_a \sin\left(\tan^{-1}\left(\sqrt{\tan^2(\theta_{xd_{1j}}-\Theta_{x_1})+\tan^2(\theta_{yd_{1j}}-\Theta_{y_1})}\right)\right)
				\sin(\phi_{u_{1j}}))}{2}\right)} 
		{N_q\sin\left(\frac{k d_a \sin\left(\tan^{-1}\left(\sqrt{\tan^2(\theta_{xd_{1j}}-\Theta_{x_1})+\tan^2(\theta_{yd_{1j}}-\Theta_{y_1})}\right)\right)
				\sin(\phi_{u_{1j}})}{2}\right)}\right)^2
	\end{align}
	\hrulefill
\end{figure*}
Also parameter $\phi_{u_{1j}}$ is the Roll angle of antenna $A_{u_{1j}}$ relative to $A_{\mathcal{S}_{11}}$, which is important in interference analysis.
For both intra-cell and inter-cell interference, the side-lobes generally cause interference. Unlike the main lobe, as shown in Fig. \ref{nv4}, the side-lobes are functions of $\phi$. 
%

{\bf Remark 1.} {\it Equation \eqref{sd1} is a function of RVs $\Theta_{x_1}$, $\Theta_{y_1}$, $\theta_{yd_{1j}}$, and $\theta_{xd_{1j}}$. However, the RVs $\theta_{x_1}$, and $\theta_{y_1}$ are caused by the UAV's vibrations and have a coherence time of less than a second. Moreover, the parameters $\theta_{yd_{1j}}$, and $\theta_{xd_{1j}}$ are caused by the distribution of SBSs and NFPs in the 3D space, which usually change for a longer period of time.}

\begin{figure*}
	\centering
	\subfloat[] {\includegraphics[width=3.5 in]{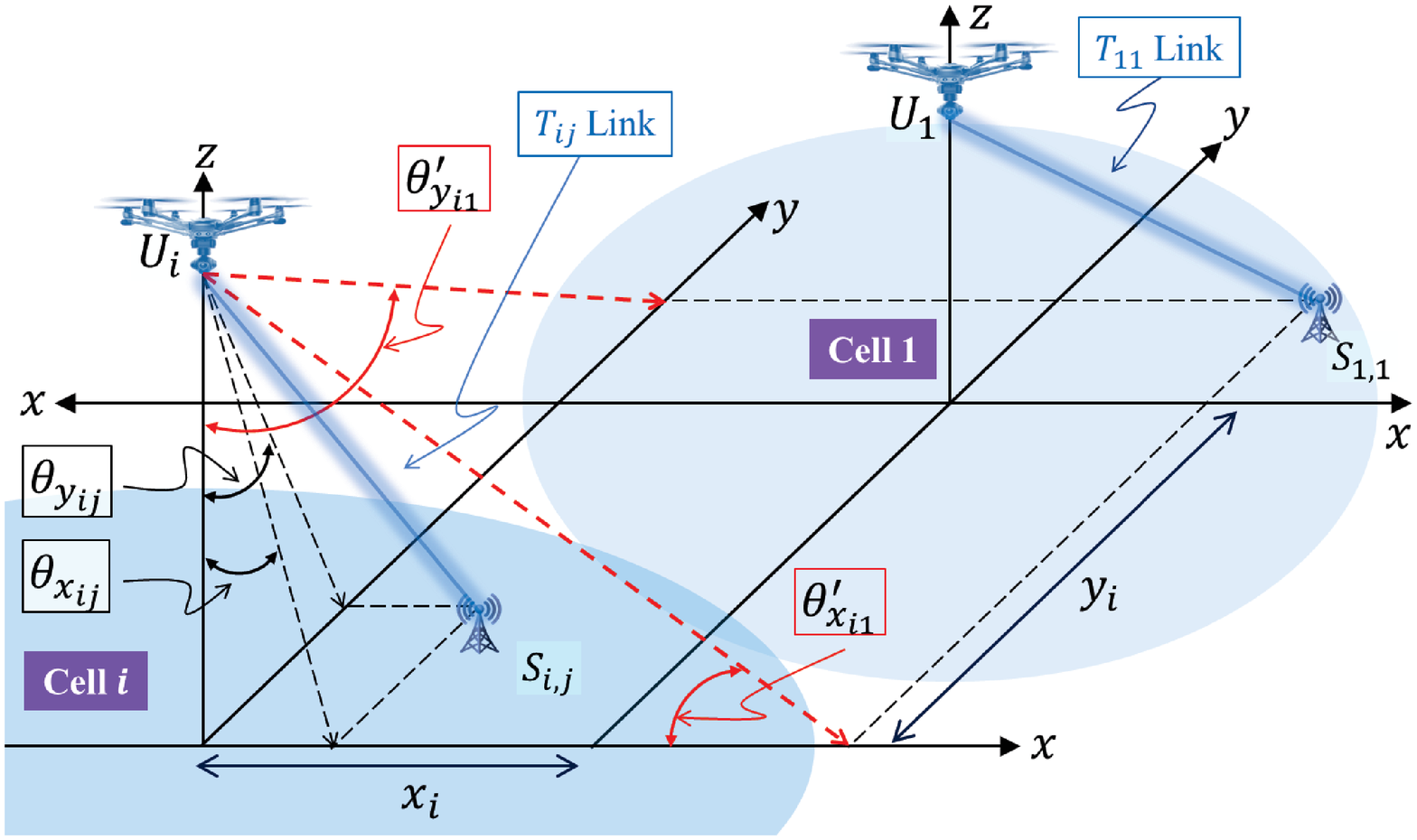}
		\label{xx1}
	}
	\hfill
	\subfloat[] {\includegraphics[width=3.5 in]{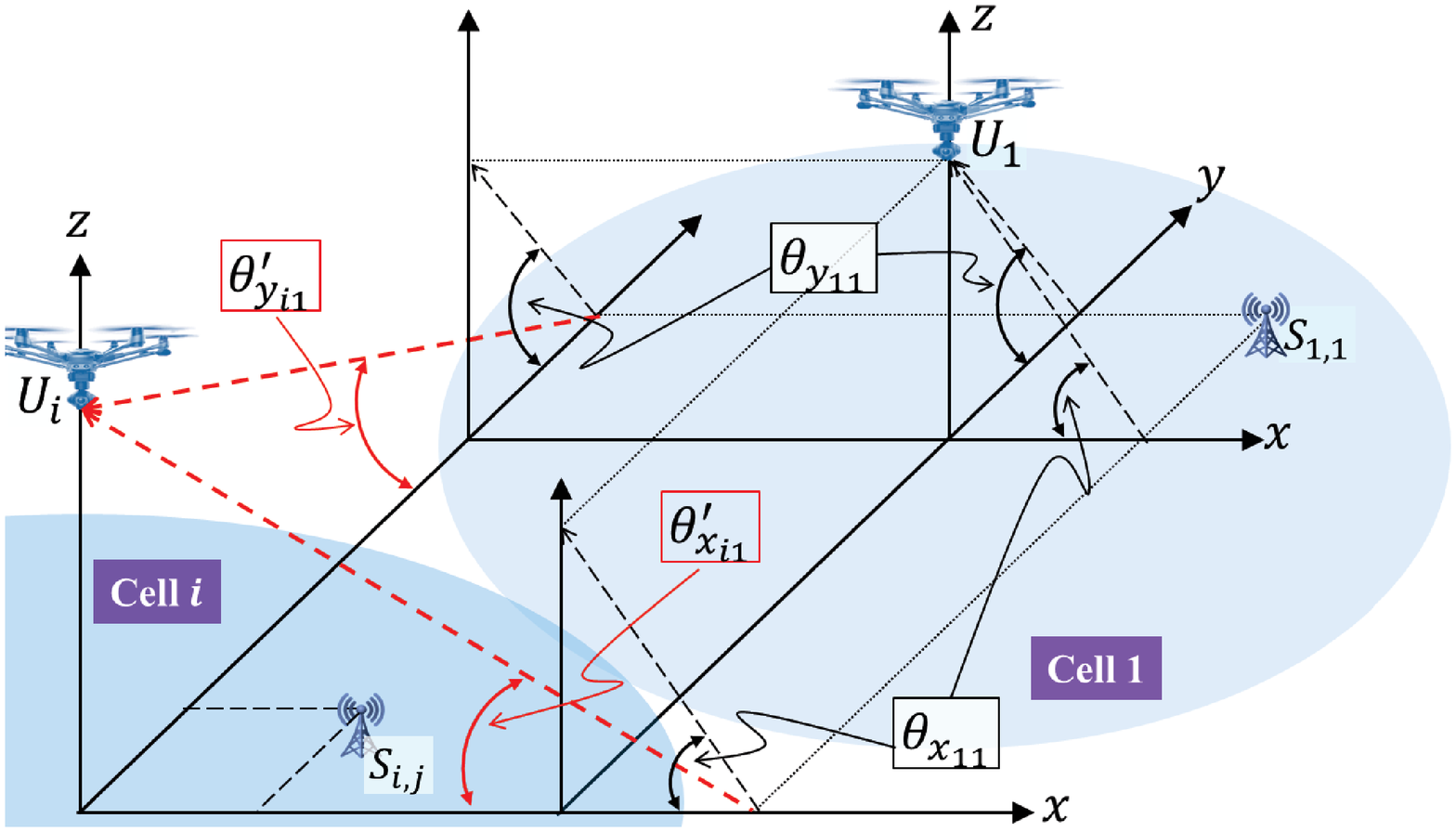}
		\label{xx2}
	}
	\caption{Characterization the interference caused by $T_{ij}$ link on the considered $T_{11}$ link.
		This interference depends on two spatial angles
		$\theta'_{i,j1}=[\theta_{xd_{i,j1}}+\Theta_{x_i},\theta_{yd_{i,j1}}+\Theta_{y_i}]$, 
		$\theta''_{i1}=[\theta_{xd_{i1}},\theta_{yd_{i1}}]$, where
		$\theta_{xd_{i,j1}}=(\theta'_{x_{i1}}-\theta_{x_{ij}})$,
		$\theta_{yd_{i,j1}}=(\theta'_{y_{i1}}-\theta_{y_{ij}})$,
		$\theta_{xd_{i1}}=(\theta'_{x_{i1}}-\theta_{x_{11}})$, and
		$\theta_{yd_{i1}}=(\theta'_{y_{i1}}-\theta_{y_{11}})$. 
		(a) Characterization of parameter $\theta'_{i,j1}$, and (b) characterization of parameter $\theta''_{i1}$.}
	\label{xx3}
\end{figure*}

\subsubsection{Inter-cell Interference}
The inter-cell interference is caused by the interference of $T_{ij}$ links on the considered $T_{11}$ link when $i\neq1$ and $j\in\{1,...,J\}$. Therefore, the inter-cell interference will be given by:
\begin{align}
	\label{n2}
	\mathbb{I}_\text{inter}=\sum_{i=2}^{I} \sum_{j=1}^J \mathbb{I}_{ij},
\end{align}
where $I$ is the number of NFPs and 
$\mathbb{I}_{ij}$ is the interference caused by a link $T_{ij}$.
As shown in Fig. \ref{xx3}, $\mathbb{I}_{ij}$ depends on the angle between the direction of $A_{u_{ij}}$ with respect to $\mathcal{S}_{11}$ denoted by $\theta'_{i,j1}=[\theta_{xd_{i,j1}}+\Theta_{x_i},\theta_{yd_{i,j1}}+\Theta_{y_i}]$, 
the angle between the direction of $A_{\mathcal{S}_{11}}$ with respect to $U_i$ denoted by $\theta''_{i,j}=[\theta_{xd_{i,j}},\theta_{yd_{i,j}}]$, and also the distance between $U_i$ and $\mathcal{S}_{11}$ denoted by $L_{i1}$. 
The position of $U_1$ is characterized as $[0,0,h_1]$ in the considered Cartesian coordinate system $[x,y,z]\in\mathbb{R}^{1\times3}$, and the position of $U_i$ for $i\in\{2,...,I\}$ are characterized as $[x_i,y_i,h_i]$. Therefore, $L_{i1}$ is obtained as
\begin{align}
	\label{c1}
	L_{i1} = \sqrt{(x_i^2+h_1 \tan(\theta_{x_{11}}))^2   +    (y_i^2+h_1 \tan(\theta_{y_{11}}))^2 + h_i^2}.
\end{align}
Also it can be easily shown that 
$\theta_{xd_{i,j1}}=(\theta'_{x_{i1}}-\theta_{x_{ij}})$,
$\theta_{yd_{i,j1}}=(\theta'_{y_{i1}}-\theta_{y_{ij}})$,
$\theta_{xd_{i1}}=(\theta'_{x_{i1}}-\theta_{x_{11}})$, and
$\theta_{yd_{i1}}=(\theta'_{y_{i1}}-\theta_{y_{11}})$, where
\begin{align}
	\theta'_{x_{i1}} &= \tan^{-1}\left( \frac{h_i}{x_i + h_1 \tan(\theta_{x_{11}})} \right) \nonumber \\
	\theta'_{y_{i1}} &= \tan^{-1}\left( \frac{h_i}{y_i + h_1 \tan(\theta_{y_{11}})} \right)
\end{align}
Based on this, $\mathbb{I}_{ij}$ is defined in \eqref{sd2}.
\begin{figure*}[!t]
	\normalsize
	\begin{align}
		\label{sd2}
		&\mathbb{I}_{ij} = P_t |h_{L_{i1}}|^2  G_0(N_s) G_0(N_u) \nonumber \\
		&~~~~\times G_e\left(\tan^{-1}\left(\sqrt{\tan^2(\theta_{xd_{i,j}})+\tan^2(\theta_{yd_{i,j}})}\right)\right) 
		G_e\left(\tan^{-1}\left(\sqrt{\tan^2(\theta'_{x_{i1}}-\theta_{x_{ij}} + \Theta_{x_i})+\tan^2(\theta'_{y_{i1}}-\theta_{y_{ij}}+\Theta_{y_i})}\right)\right) 
		\nonumber \\
		&~~~~\times\left( \frac{\sin\left(\frac{N_q (k d_a \sin\left(\tan^{-1}\left(\sqrt{\tan^2(\theta_{xd_{i,j}})+\tan^2(\theta_{yd_{i,j}})}\right)\right)
				\cos(\phi_{\mathcal{S}_{11}}))}{2}\right)} 
		{N_q\sin\left(\frac{k d_a \sin\left(\tan^{-1}\left(\sqrt{\tan^2(\theta_{xd_{i,j}})+\tan^2(\theta_{yd_{i,j}})}\right)\right)
				\cos(\phi_{\mathcal{S}_{11}})}{2}\right)}
		 \frac{\sin\left(\frac{N_q (k d_a \sin\left(\tan^{-1}\left(\sqrt{\tan^2(\theta_{xd_{i,j}})+\tan^2(\theta_{yd_{i,j}})}\right)\right)
				\sin(\phi_{\mathcal{S}_{11}}))}{2}\right)} 
		{N_q\sin\left(\frac{k d_a \sin\left(\tan^{-1}\left(\sqrt{\tan^2(\theta_{xd_{i,j}})+\tan^2(\theta_{yd_{i,j}})}\right)\right)
				\sin(\phi_{\mathcal{S}_{11}})}{2}\right)}\right)^2 \nonumber \\
		&~~~~\times\left( \frac{\sin\left(\frac{N_q (k d_a \sin\left(\tan^{-1}\left(\sqrt{\tan^2(\theta'_{x_{i1}}-\theta_{x_{ij}} + \Theta_{x_i})+\tan^2(\theta'_{y_{i1}}-\theta_{y_{ij}}+\Theta_{y_i})}\right)\right)
				\cos(\phi_{u_{ij}}))}{2}\right)} 
		{N_q\sin\left(\frac{k d_a \sin\left(\tan^{-1}\left(\sqrt{\tan^2(\theta'_{x_{i1}}-\theta_{x_{ij}} + \Theta_{x_i})+\tan^2(\theta'_{y_{i1}}-\theta_{y_{ij}}+\Theta_{y_i})}\right)\right)
				\cos(\phi_{u_{ij}})}{2}\right)}
		\right. \nonumber \\
		&~~~~\times \left. \frac{\sin\left(\frac{N_q (k d_a \sin\left(\tan^{-1}\left(\sqrt{\tan^2(\theta'_{x_{i1}}-\theta_{x_{ij}} + \Theta_{x_i})+\tan^2(\theta'_{y_{i1}}-\theta_{y_{ij}}+\Theta_{y_i})}\right)\right)
				\sin(\phi_{u_{ij}}))}{2}\right)} 
		{N_q\sin\left(\frac{k d_a \sin\left(\tan^{-1}\left(\sqrt{\tan^2(\theta'_{x_{i1}}-\theta_{x_{ij}} + \Theta_{x_i})+\tan^2(\theta'_{y_{i1}}-\theta_{y_{ij}}+\Theta_{y_i})}\right)\right)
				\sin(\phi_{u_{ij}})}{2}\right)}\right)^2 \times P_\text{LoS}(\theta_{\text{elev},i}), \\
		&\text{where}\nonumber \\	
		& P_\textrm{LoS}(\theta_{\text{elev},i}) = \frac{1}
		{1+\alpha \exp\left(-\beta\left[ 
			\frac{180}{\pi}\tan^{-1}\left(\sqrt{\left( \frac{h_i}{x_i + h_1 \tan(\theta_{x_{11}})} \right)^2
				+\left( \frac{h_i}{y_i + h_1 \tan(\theta_{y_{11}})} \right)^2}\right)
			-\alpha\right]\right)}	.
	\end{align}
	\hrulefill
\end{figure*}
%
%

{\bf Remark 2.} {\it In order to compute inter-cell interference from \eqref{sd2}, for any distribution of SBSs and $U_i$, we calculate parameters $\theta'_{w_{ij}}$, $\theta_{w_{ij}}$, and $L_{i1}$ for $w\in\{x,y\}$, $I\in\{1,...,I\}$, and $j\in\{1,...,J\}$ based on the positions of SBSs and $U_i$ in the considered Cartesian coordinate system. For any change in the distribution of SBSs and/or the position of hovering $U_i$, we must recalculate and update the indicated parameters.
However, the parameters $\Theta_{x_i}$ and $\Theta_{y_i}$ are caused by the UAVs' angular vibrations and have a coherence time of less than a second, and act as rapid RVs in the analyses compared to the RVs $\theta'_{w_{ij}}$, $\theta_{w_{ij}}$, and $L_{i1}L_{i1}$.}

By substituting \eqref{p3}, \eqref{sd1} and \eqref{sd2}, in \eqref{p1}, the instantaneous value of $\gamma_{11}$ is calculated, which is a complex expression of the channel parameters. Before analytical derivations which is provided in Section IV, it is necessary to have a preliminary view of the effect of the parameters on the $\gamma_{11}$ and model it more simple.
For this aim, in the next subsection, by providing simulation, we will study the inter-cell and intra-cell interference, which leads to important results. The obtained results will let us to simplify $\gamma_{11}$ and prepare the groundwork for the comprehensive analysis of the system in the next section.

\begin{table}
	\caption{Parameter values for simulations of single relay system.} 
	\centering 
	\begin{tabular}{| l |c || l| c|} 
		\hline\hline 
		{\bf Parameters} & {\bf Values} &
		{\bf Parameters} & {\bf Values}  \\ [.5ex] 
		\hline\hline 
		%
		$P_{t} $ & 10 mW &
		$\gamma_\text{th}$ & 9 dB\\ \hline
		$N_{s}$ & 15 &
		$N_{u}$ & 5-25 \\ \hline
		$f_c$ &  95 GHz &
		%
			$J$ & 10\\ \hline
			$R$&   $400$ m  &
			$T$ & $20^o$C\\ \hline
			BW & 3 GHz &
			$\sigma_\theta$ & $0.5^o-1.5^o$ \\ \hline
			$h_i$ &  50-100 m &  $d_{x}=d_{y}$ & $\lambda/2$ \\ 
			\hline \hline              
	\end{tabular}
	\label{I2} 
\end{table}

\subsection{Simulation Results}

The values used for the simulations are provided in Table \ref{I2}. 
In our simulation setting, we randomly deploy (position of each SBS is a uniform random variable on $x-y$ plane) 100 SBSs are randomly distributed in an area on $x-y$ plane with a radius of 1.5\,km. Then, ten UAVs are placed at a height of 80-110 meter from the ground level in such a way that each UAV is connected to the nearest ten SBSs. An SBS is chosen randomly and we call it $\mathcal{S}_{11}$ and the UAV connected to it is called $U_1$. 
According to the position of the other SBSs and $U_i$, we calculate the average value of the parameters related to angle and distance. Then, $5\times10^7$ random independent values of RVs $\Theta_{x_i}$ and $\Theta_{y_i}$ are generated related to the UAVs' vibrations. After that, using \eqref{n1} and \eqref{sd1}, $5\times10^7$ independent values of intra-cell interference and using \eqref{n2} and \eqref{sd2}, $5\times10^7$ independent values of inter-cell interference are produced.
Finally, by averaging over $5\times10^7$ independent runs, the average values of inter-cell and intra-cell interference are obtained.

\begin{figure}
	\begin{center}
		\includegraphics[width=3.4 in]{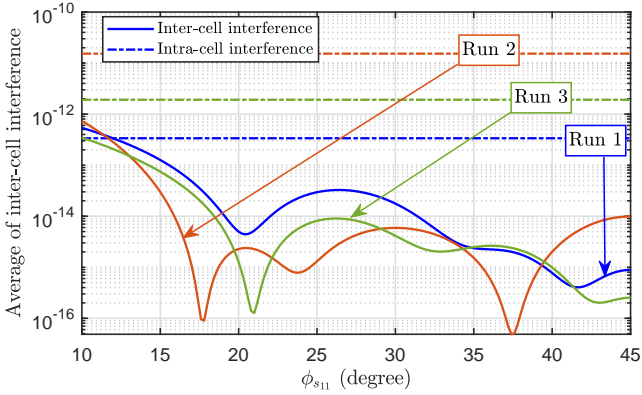}
		\caption{Comparison of inter- and intra-cell interference versus $\phi_{\mathcal{S}_{11}}$ for three independent random distributions of SBSs.}
		\label{vv1}
	\end{center}
\end{figure}

\begin{figure}
	\begin{center}
		\includegraphics[width=3.4 in]{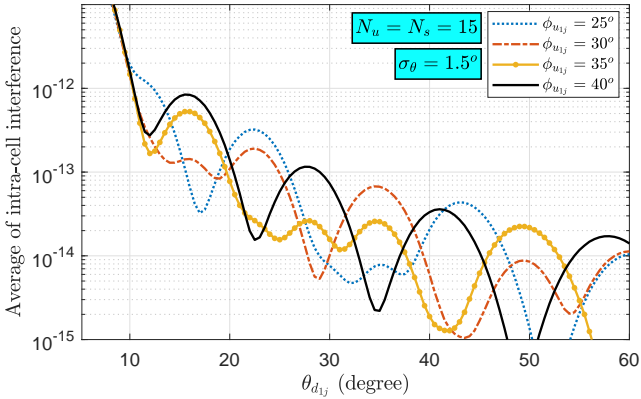}
		\caption{Comparison of intra-cell interference due to $T_{1j}$ link versus $\theta_{d_{1j}}$ for different values of $\phi_{u_{1j}}$.}
		\label{vv2}
	\end{center}
\end{figure}

The papremter $\phi$ is one of the important parameters for the interference analysis. To show its importance, in Fig. \ref{vv1}, the inter-cell and intra-cell interference are plotted for $N_u=15$. The results of Fig. \ref{vv1} are provided for three independent random distributions of SBSs. For each independent distribution, similar to the steps mentioned above, the inter-cell and intra-cell interference are plotted, separately, and indicated by ``Run $n$'' in Fig. \ref{vv2} for $n\in\{1,2,3\}$.
As can be observed from the results of Fig. \ref{vv1}, with any change in the distribution of SBSs and NFPs, the inter-cell and intra-cell interference changes. A very important point is that with the change of $\phi_{\mathcal{S}_{11}}$, the inter-cell interference changes significantly. Since $\phi_{\mathcal{S}_{11}}$ is a tunable parameter, for any distribution of SBSs and NFPs, the parameter $\phi_{\mathcal{S}_{11}}$ can be adjusted in such a way that the inter-cell interference falls below the noise level and becomes almost negligible. For example, based on the results of Fig. \ref{vv1}, for {\bf Run 1}, the inter-cell interference falls bellow $10^{-15}$ for $40^o<\phi_{\mathcal{S}_{11}}<45$, while for {\bf Run 2}, the inter-cell interference reaches less than $10^{-16}$ for $\phi_{\mathcal{S}_{11}}=35^o$, which is much lower than the noise level and/or intra-cell interference.
However, changing  $\phi_{\mathcal{S}_{11}}$ has almost no effect on the intra-cell interference. To find out the reason for this, it should be noted that $\phi_{\mathcal{S}_{11}}$ is related to the antenna of $\mathcal{S}_{11}$, which is directed towards $U_1$. Therefore, the main lobe of $A_{\mathcal{S}_{11}}$ is towards $A_{u_{1j}}$, and as can be observed from the results of Fig. \ref{nv6}, 
changing parameter $\phi_{\mathcal{S}_{11}}$ has almost no effect on the main lobe antenna pattern. As a result, the intra-cell interference is constant with respect to the change of $\phi_{\mathcal{S}_{11}}$.
Due to the importance of the obtained results, in the following remarks, a summary of the obtained results is provided, which will be used in the next section for the analysis and parameter system design.

{\bf Remark 3.} {\it For any distribution of SBSs and NFPs, the tunable parameter  $\phi_{\mathcal{S}_{11}}$ can be set in such a way that the inter-cell interference falls below the noise level and thus, inter-cell interference has a negligible destructive effect on the considered system performance.}

{\bf Remark 4.} {\it Since the antenna pattern in the main lobe is constant with respect to $\phi$, the parameter $\phi_{\mathcal{S}_{11}}$ has almost no effect on the intra-cell interference.}

%

%

Unlike inter-cell interference, intra-cell interference is the main factor in reducing the quality of the considered system, and cannot be ignored due to following reasons: (i) as discussed, the main lobe of $A_{\mathcal{S}_{11}}$ is set toward $A_{u_{11}}$ antenna mounted on $U_1$, and unintentionally, it is also set toward the $A_{u_{1j}}$ antennas.
(ii) The intra-cell interference caused by $A_{u_{1j}}$ has a smaller distance $A_{\mathcal{S}_{11}}$ than the inter-cell interference caused by $A_{u_{ij}}$, and thus, it has less attenuation.
(iii) Most importantly, compared to $A_{u_{ij}}$ antennas, with more probability, $A_{u_{1j}}$ antennas are placed at a smaller spatial angle than $A_{\mathcal{S}_{11}}$.
For better understanding, according to \eqref{sd2}, the intra-cell interference is caused by the interference between the main lobe of $A_{\mathcal{S}_{11}}$ and the side-lobe of the $A_{u_{1j}}$ antennas. For interference limited scenarios (i.e., SINR tends to signal-to-interference ratio (SIR)), 
since the main lobe of $A_{\mathcal{S}_{11}}$ antenna simultaneously boosts the main signal and intra-cell interference, changing the antenna pattern of $A_{\mathcal{S}_{11}}$ has no effect on the SIR. Therefore, the only option to reduce the intra-cell interference is to reduce the interference caused by the side-lobes of $A_{u_{1j}}$, which is a function of the antenna pattern as well as the spatial angle $\theta_{d_{1j}}$ and $\phi_{u_{1j}}$a.
In particular, the spatial angle $\theta_{d_{1j}}$ is a function of the distribution of $\mathcal{S}_{1j}$ with respect to the $U_1$, which is a random distribution. 
For better understanding, in Fig. \ref{vv2}, the intra-cell interference is drawn versus $\theta_{d_{1j}}$ and different values of $\phi_{u_{1j}}$.
As we expected, by increasing $\theta_{d_{1j}}$, the antenna side-lobes decreases and, as a result, the intra-cell interference decreases. However, as can be seen, the interference does not decrease linearly by increasing $\theta_{d_{1j}}$ and depends on the parameter $\phi_{u_{1j}}$.
An important point is that for small values of $\theta_{d_{1j}}$, the intra-cell interference is approximately caused by the main lobe of the antennas, and in this case, the communication link will definitely be interrupted. This is related to the case when the distance of one of the $\mathcal{S}_{1j}$ is very close to $\mathcal{S}_{11}$ (in the order of several meters). In this case, as predicted in 5G, two close frounthaul links should be integrated together, which is easily possible due to their very small distance.

{\bf Remark 5.} {\it According to the above results, a minimum angle $\alpha_c$ is defined for the $\mathcal{S}_{1j}$ that are close to each other. If the spatial angle $\theta_{d_{1j}}$ between $T_{11}$ and $T_{1j}$ links is less than $\alpha_c$, both frounthaul links should be integrated according to the protocols defined in 5G.}

\section{Outage Probability Analysis}

In this section, we use the results obtained in Section III, and we analyze the performance of the considered system in terms of outage probability.

\subsection{Outage Probability Analysis}
Outage probability of $T_{11}$ link (i.e., the probability that $\gamma_{11}$ falls below a threshold $\gamma_\text{th}$) is obtained as follows:
\begin{align}
	\mathbb{P}_\text{out}= \int_0^{\gamma_\text{th}} f_{\gamma_{11}}(\gamma_{11}) \text{d}\gamma_{11}.
\end{align}
Therefore, to analyze the considered system, the first step is to know the distribution function of SINR $\gamma_{11}$. 
In the following theorem, the PDF of $\gamma_{11}$ is provided.

{\bf Theorem 1.}
{\it  For any random distribution of SBSs, the PDF of $\gamma_{11}$ for the worst case scenario is obtained as:}
\begin{align}
	\label{c7}
	&f_{\gamma_{11}|\theta_{d_{1j}}}(\gamma_{11}) = \frac{1}{\sqrt{\beta_{qx}\beta_{qy}}} \\
	&
	\left[ \sum_{j=1}^J \sum_{m=0}^M A_m A''_m e^{-\frac{(\theta_{d_{1j}}-A'_m)^2}{w_m^2} } + N'_0 \right]^{\frac{\beta_{qx}+\beta_{qy}}{2\beta_{qx}\beta_{qy}} }
	h_p^{\frac{\beta_{qx}+\beta_{qy}}{2\beta_{qx}\beta_{qy}} -1 } \nonumber \\ 
	&I_0\left(\beta_{qxy}
	\ln\left(\left[ \sum_{j=1}^J \sum_{m=0}^M A_m A''_m e^{-\frac{(\theta_{d_{1j}}-A'_m)^2}{w_m^2} } + N'_0 \right]\gamma_{11}\right) \right) \nonumber
\end{align}
{\it where $\beta_{uw}= \frac{\sqrt{2}\sigma_{\theta_w}^2}{w_0}$, $\beta_{qxy}=  \frac{\beta_{qy}-\beta_{qx}}{2\beta_{qx}\beta_{qy}}$, 
	$N'_0= \frac{\sigma^2_{N}}{P_t |h_{L_{11}}|^2 G_0(N_s) G_0(N_u)}$ is the normalized thermal noise,
	\begin{align}
		\left\{ \!\!\!\!\! \! \!
		\begin{array}{rl}
			& w_m = \frac{1}{N_s}~~\&~A'_m=0,~~~~~~~~~~~~~~~~~~~~m=0 \\
			& w_m = \frac{1}{2N_s}~\&~A'_m=\sin^{-1}\left( \frac{2m+1}{N_u}\right),~~~~m\neq0 \\
		\end{array} \right. \nonumber
	\end{align}
	$$\left\{ \!\!\!\!\! \! \!
	\begin{array}{rl}
		& A''_m=1,~~\text{for}~~\frac{m}{N_u}\leq1 \\
		& A''_m=0,~~\text{for}~~\frac{m}{N_u}>1 \\
	\end{array} \right. $$
	and $M=6$, $A_0=1, A_1=0.05, A_2=0.020, A_3=0.011, A_4=0.006, A_5=0.004$, $A_6=0.003$.}
%
\begin{IEEEproof}
	Please refer to Appendix \ref{AppA}. 
\end{IEEEproof} 

Notice that, as can be seen from \eqref{sd1} and \eqref{sd2}, $\gamma_{11}$ is a complex function of channel RVs, that include the location of SBSs relative to each other as well as relative to the UAV connected to it, and the orientation fluctuations of the directional antennas mounted on hovering UAVs.
Therefore, it seems impossible to obtain the exact distribution function for $\gamma_{11}$. Using the results obtained in the Section III, and using some derivations, the distribution function of $\gamma_{11}$ for the worst-case is provided in Theorem 1. In network design, operators usually design the system for the worst-case scenario or consider a guard band for analysis. 
In order to derive the PDF of $\gamma_{11}$ in \eqref{c7}, we provide and use the approximated antenna pattern in \eqref{c5} as a function of the sum of several Gaussian functions.
As we see from Fig. \ref{nv5}, the approximated model provided in \eqref{c5} models well the main lobe along with the maximum amount of side-lobes of real antenna pattern.
We use \eqref{c5} to model the interference, and since the actual value of the interference is equal to or less than the interference obtained from \eqref{c5}, the interference obtained from \eqref{c5} is for the worst scenario.

{\bf Theorem 2.}
{\it  For any random distribution of SBSs, the outage probability for the worst case scenario is derived \eqref{c8}.}
\begin{figure*}[!t]
	\normalsize
	\begin{align}
		\label{c8}
		&\resizebox{0.99\hsize}{!}{$
			\mathbb{P}_{\text{out}|\theta_{d_{1j}}} = 1 -
			Q\left( \mathbb{B}_1 \sqrt{\frac{-2\ln\left(\left[ \sum_{j=1}^J \sum_{m=0}^M A_m A''_m e^{-\frac{(\theta_{d_{1j}}-A'_m)^2}{w_m^2} } + N'_0 \right]\gamma_{th}\right)}{\beta_{ux}+\beta_{uy}}}  ,  \mathbb{B}_2 \sqrt{\frac{-2\ln\left(\left[ \sum_{j=1}^J \sum_{m=0}^M A_m A''_m e^{-\frac{(\theta_{d_{1j}}-A'_m)^2}{w_m^2} } + N'_0 \right]\gamma_{th}\right)}{\beta_{ux}+\beta_{uy}}}  \right)$} \nonumber \\ 
		&\resizebox{0.9\hsize}{!}{$
			+  Q\left( \mathbb{B}_2 \sqrt{\frac{-2\ln\left(\left[ \sum_{j=1}^J \sum_{m=0}^M A_m A''_m e^{-\frac{(\theta_{d_{1j}}-A'_m)^2}{w_m^2} } + N'_0 \right]\gamma_{th}\right)}{\beta_{ux}+\beta_{uy}}}  ,  \mathbb{B}_1 \sqrt{\frac{-2\ln\left(\left[ \sum_{j=1}^J \sum_{m=0}^M A_m A''_m e^{-\frac{(\theta_{d_{1j}}-A'_m)^2}{w_m^2} } + N'_0 \right]\gamma_{th}\right)}{\beta_{ux}+\beta_{uy}}}  \right)$}.
	\end{align}
	\hrulefill
\end{figure*}
{\it In \eqref{c8}, $Q(a,b)$ is the Marcum {\it Q}-function, $\gamma_{th}$ is the SINR threshold, and} 
\begin{align}
		\left\{ \!\!\!\!\! \! \!
		\begin{array}{rl}
			&\mathbb{B}_1 = \frac{\sqrt{1-T_q^4}}{2T_q} \sqrt{\frac{1+T_q}{1-T_q}},\\
			&\mathbb{B}_2 = \mathbb{B}_1 \frac{1-T_q}{1+T_q},   ~~~ 
			T_q  = \frac{\sigma_{\theta_\text{min}}}{\sigma_{\theta_\text{max}}}, \\
			&\sigma_{\theta_\text{max}} = \text{max}\{\sigma_{\theta_{x}},\sigma_{\theta_{y}}\},~~~~
			\sigma_{\theta_\text{min}} = \text{min}\{\sigma_{\theta_{x}},\sigma_{\theta_{y}}\}
		\end{array} \right. \nonumber
\end{align}
\begin{IEEEproof}
	Please refer to Appendix \ref{AppB}. 
\end{IEEEproof}

The outage probability obtained in Theorem 2 is based on the distribution function obtained in Theorem 1 and therefore calculates the upper bound of outage probability. Another point is that the outage probability provided in Theorem 2 is conditional on knowing $\theta_{d_{1j}}$.
The parameter $\theta_{d_{1j}}$ are function of the position of the $\mathcal{S}_{1j}$ relative to the connected UAV, which are random variables. Therefore, in order to obtain the average outage probability, it is necessary to have the distribution function of $\theta_{d_{1j}}$, which is analyzed in the following two propositions.

{\bf Proposition 1.}
{\it  The PDF of RV $\theta_{11}$ is obtained as:}
\begin{align}
	\label{pk7}
	f_{\theta_{{11}}} (\theta_{11}) = \frac{h_1}{R \cos^2(\theta_{{11}})}  ~~\text{for}~~ 0 <\theta_{11}<\tan^{-1}\left(\frac{R}{h_1}\right). 
\end{align}
\begin{IEEEproof}
	Please refer to Appendix \ref{AppC}. 
\end{IEEEproof}

\begin{figure}
	\centering
	\subfloat[] {\includegraphics[width=3.3 in]{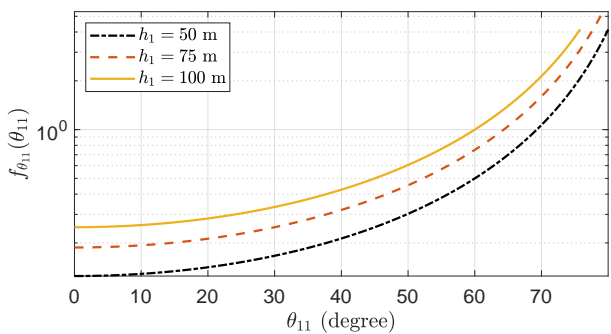}
		\label{b1}
	}
	\hfill
	\subfloat[] {\includegraphics[width=3.3 in]{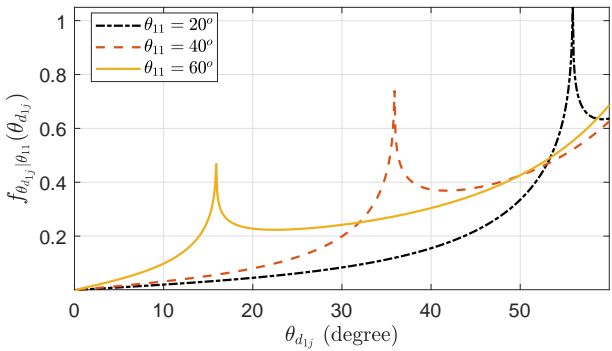}
		\label{b2}
	}
	\caption{(a) The distribution function of $\theta_{x_{11}}$ for different $U_1$'s flight heights and $R=400$m.
		(b) The distribution function of $\theta_{d_{1j}}$ for different values of $\theta_{x_{11}}$.}
	\label{b3}
\end{figure}

The distribution of $\theta_{x_{11}}$ which determines the position of $\mathcal{S}_{11}$ is very important for performance analysis because the intra-cell interference depends on the position $\mathcal{S}_{11}$.
In Proposition 1, the distribution function of $\theta_{x_{11}}$ is provided, which is a function of the UAV's flight height $h_1$ and the UAV's coverage radius $R$. In Fig. \ref{b1}, the distribution function of $\theta_{x_{11}}$ is plotted for different flight heights and $R=400$m. 
The results of Fig. \ref{b1} show that $f_{\theta_{11}}(\theta_{11})$ is an ascending function of $\theta_{11}$.
It means that with a low probability, $S_{11}$ is located at a small spacial angle with respect to the $U_1$ and as the height of $U1$ decreases, $f_{\theta_{11}}$ decreases for lower values of $\theta_{11}$, which is not desirable.


{\bf Proposition 2.}
{\it  The PDF and CDF of $\theta_{d_{1j}}$ are given by:}
\begin{align}
	\label{nk6}
	f_{\theta_{d_{1j}}}(\theta_{d_{1j}}) = \int f_{\theta_{d_{1j}}  | \theta_{x_{1j}} }(\theta_{d_{1j}}) 
	f_{ \theta_{x_{1j}} }(\theta_{x_{1j}}) \text{d} \theta_{x_{1j}}.
\end{align}
\begin{align}
	\label{nk7}
	F_{\theta_{d_{1j}}}(\theta_{d_{1j}}) = \int F_{\theta_{d_{1j}}  | \theta_{x_{1j}} }(\theta_{d_{1j}}) 
	f_{ \theta_{x_{1j}} }(\theta_{x_{1j}}) \text{d} \theta_{x_{1j}},
\end{align} 
{\it where}
\begin{align}
	\label{nk3}
	& f_{\theta_{d_{1j}} | \theta_{x_{1j}}} (\theta_{d_{1j}}) 
	= \frac{h_1}{\sqrt{R^2-h_1^2\tan^2(\theta_{x_{1j}})}} \nonumber \\
	&\times\frac{\sin(\theta_{d_{1j}})}
	{\cos^3(\theta_{d_{1j}}) \sqrt{\tan^2(\theta_{d_{1j}})  -  \tan^2(\theta_{x_{1j}}-\theta_{x_{11}}) }}
\end{align}
\begin{align}
	\label{nk8}
	&F_{\theta_{d_{1j}} | \theta_{x_{1j}}} (\theta_{d_{1j}}) = \\
	&\frac{h_1}{\sqrt{R^2-h_1^2\tan^2(\theta_{x_{1j}})}}
	\sqrt{\tan^2(\theta_{d_{1j}})  -  \tan^2(\theta_{x_{1j}}-\theta_{x_{11}}) }, \nonumber
\end{align}
\begin{align}
	\label{nk9}
	\resizebox{0.99\hsize}{!}{$
		f_{ \theta_{x_{1j}} }(\theta_{x_{1j}}) = 
		\frac{h_1}{R \cos^2(\theta_{x_{1j}})},~~~
		-\tan^{-1}\left(\frac{R}{h_1}\right)<\theta_{x_{1j}}<\tan^{-1}\left(\frac{R}{h_1}\right).
		$}
\end{align}

\begin{IEEEproof}
	Please refer to Appendix \ref{AppD}. 
\end{IEEEproof}

Note that $\theta_{d_{1j}}$ is a function of $\theta_{x_{1j}}$ and $\theta_{y_{1j}}$. In order to calculate the average outage probability, we need to know the PDF and CDF of $\theta_{d_{1j}}$, which are derived in Proposition 2. The important point is that $\theta_{d_{1j}}$ is also a function of $\theta_{11}$ which is modeled in proposition 1.
In Fig. \ref{b2}, the distribution function of $\theta_{d_{1j}}$ is plotted for different values of $\theta_{11}$. As mentioned in the previous section, the larger $\theta_{d_{1j}}$ is, the less interference it creates, and as a result, the system performance improves.
The results of Fig. \ref{b2} show that the smaller $\theta_{11}$ is, the probability of having large values of $\theta_{d_{1j}}$ increases, and as a result, we expect that the interference decreases and the system performance improves. 
On the other hand, according to the Proposition 1, as the flight height increases, the probability of having a smaller $\theta_{11}$ increases, which leads to a decrease in the average outage probability. The obtained results are summarized in the remark below.

{\bf Remark 6.} {\it Based on the results obtained from propositions 1 and 2, as the flight height increases, the probability of smaller values of $\theta_{d11}$ decreases, and as a result, the interference decreases and the system performance improves.}
{\bf Theorem 3.}
{\it  The average outage probability for the worst case scenario is derived as:}
\begin{align}
	\label{bn1}
	\mathbb{P}_{\text{out}} = \int \mathbb{P}_{\text{out}|\theta_{x_{1j}}} f_{x_{1j}}(x_{1j}) \text{d} \theta_{x_{1j}}
\end{align}
{\it where $\mathbb{P}_{\text{out}|\theta_{x_{1j}}}$ is derived in \eqref{pn2}. In \eqref{pn2}, we have $\mathbb{G}_1=\max\{\alpha_c,|\theta_{x_{1j}}-\theta_{x_{11}}|\}$, and $\mathbb{G}_2 = (\theta_{d\text{max}}-\mathbb{G}_1)$.}
\begin{figure*}[!t]
	\normalsize
	\begin{align}
		\label{pn2}
		& \mathbb{P}_{\text{out}|\theta_{d_{1j}}} = 
		\mathbb{P}_{\text{out}|\theta_{x_{1j}}} = \sum_{d=0}^{D-1}\sum_{j=1}^{J-1}\binom{J}{j} 
		\left( \frac{h_1}{\sqrt{R^2-h_1^2\tan^2(\theta_{x_{1j}})}} \right)^{J-1}\nonumber \\
		&\resizebox{0.9\hsize}{!}{$
			\left[ 1 -
			Q\left( \mathbb{B}_1 \sqrt{\frac{-2\ln\left(\left[j \sum_{m=0}^M A_m A''_m e^{-\frac{(\frac{d\mathbb{G}_2}{D}+\mathbb{G}_1-A'_m)^2}{w_m^2} } + N'_0 \right]\gamma_{th}\right)}{\beta_{ux}+\beta_{uy}}}  ,  \mathbb{B}_2 \sqrt{\frac{-2\ln\left(\left[ j \sum_{m=0}^M A_m A''_m e^{-\frac{(\frac{d\mathbb{G}_2}{D}+\mathbb{G}_1-A'_m)^2}{w_m^2} } + N'_0 \right]\gamma_{th}\right)}{\beta_{ux}+\beta_{uy}}}  \right) \right.$} \nonumber \\ 
		&\resizebox{0.9\hsize}{!}{$  \left.
			+  Q\left( \mathbb{B}_2 \sqrt{\frac{-2\ln\left(\left[ j \sum_{m=0}^M A_m A''_m e^{-\frac{(\frac{d\mathbb{G}_2}{D}+\mathbb{G}_1-A'_m)^2}{w_m^2} } + N'_0 \right]\gamma_{th}\right)}{\beta_{ux}+\beta_{uy}}}  ,  \mathbb{B}_1 \sqrt{\frac{-2\ln\left(\left[ j \sum_{m=0}^M A_m A''_m e^{-\frac{(\frac{d\mathbb{G}_2}{D}+\mathbb{G}_1-A'_m)^2}{w_m^2} } + N'_0 \right]\gamma_{th}\right)}{\beta_{ux}+\beta_{uy}}}  \right)\right]$} \nonumber\\
		&\resizebox{0.93\hsize}{!}{$
			\left[\sqrt{\tan^2\left(\frac{(d+1)\mathbb{G}_2}{D}+\mathbb{G}_1 \right)   -  \tan^2(\theta_{x_{1j}}-\theta_{x_{11}}) } - 
			\sqrt{\tan^2\left(\frac{d\mathbb{G}_2}{D}+\mathbb{G}_1 \right)   -  \tan^2(\theta_{x_{1j}}-\theta_{x_{11}}) } \right]^{j}
			\left[1-  \sqrt{\tan^2\left(\frac{(d+1)\mathbb{G}_2}{D}+\mathbb{G}_1 \right)   -  \tan^2(\theta_{x_{1j}}-\theta_{x_{11}}) }  \right]^{J-j-1}
			$}
	\end{align}
	\hrulefill
\end{figure*}  

\begin{IEEEproof}
	Please refer to Appendix \ref{AppE}. 
\end{IEEEproof}

The number of random parameters of the considered system is at least $2(J+1)$, therefore it requires to solve a $2(J+1)$-dimensional integral, which will be very time-consuming.
To overcome this problem, in Theorem 3, the upper bound of the average outage probability of the considered system is derived, which only requires solving a one-dimensional integral and reduces the computational load, significantly.

\subsection{Numerical and Simulation Results}
Next we provide extensive simulation and numerical results in order to assess the performance of the considered system in term of outage probability. In our simulations, we follow the same path presented in the previous section for simulating interference. After obtaining $5\times10^7$ independent values of intra- and inter-cell interference, we calculate $5\times10^7$ independent values of the SINR and finally calculate the outage probability. 
This outage probability is obtained for only a random distribution of SBSs whose simulation time is in the order of minutes. To calculate the average outage probability, we simulate the outage probability for 500 random distributions of SBSs and finally calculate the average outage probability, which requires several hours of simulation time.
However, using the results of Theorem 3, the upper bound of outage probability will be computed in the order of one second.

\begin{figure}
	\begin{center}
		\includegraphics[width=3.4 in]{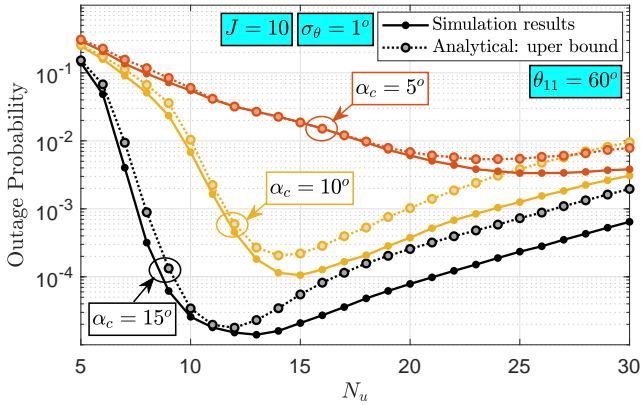}
		\caption{ Outage probability versus $N_u$ for different values of $\alpha_c$.
		}
		\label{vn1}
	\end{center}
\end{figure}

In Fig. \ref{vn1}, the analytical results obtained from Theorem 3 are compared with the outage probability results obtained from the simulations.
As we observe, the analytical results compute the upper bound of real outage probability with acceptable accuracy while it reduce the simulation time, significantly.
In Fig. \ref{vn1}, the results are obtained for different values of $\alpha_c$. 
As we expect, for the larger $\alpha_c$, the probability of having $\theta_{d_{1j}}<\alpha_c$ becomes zero and thus, the outage probability of the considered system decreases. 
Another important point is related to the antenna pattern, which is controlled by parameter $N_u$. The results of Fig. \ref{vn1} clearly show that the parameter $N_u$ has a significant effect on the system performance. 
First, by increasing $N_u$, the main received power increases and also the side-lobes become smaller and the interference decreases.
Therefore, by increasing $N_u$, the outage probability decreases. 
However, by increasing $N_u$, the beam-width decreases and the sensitivity of the antenna alignment to the UAV's vibrations increases. Therefore, the optimal selection $N_u$ is of great importance, and the provided analytical expressions help us to analyze the effect of various parameters on the system performance in a short time instead of time consuming Monte-Carlo simulations.

\begin{figure}
	\begin{center}
		\includegraphics[width=3.4 in]{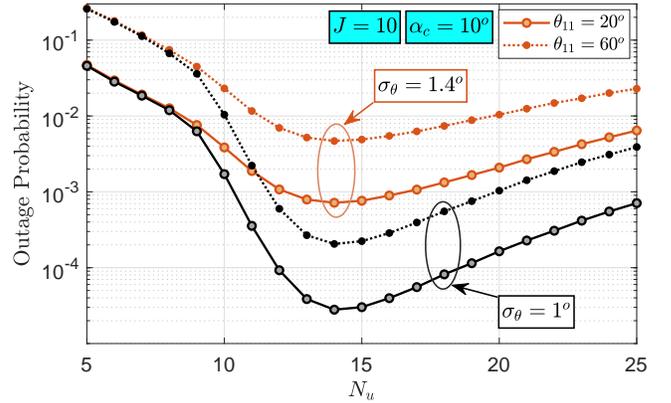}
		\caption{ Outage probability versus $N_u$ for two different valus of $\sigma_{\theta}$ and two different values of $\theta_{11}$.
		}
		\label{vn2}
	\end{center}
\end{figure}

In addition to $\alpha_c$, there are other important parameters that affect the performance of the considered system, the most important of which are $\sigma_\theta$ and $\theta_{11}$.
To get a better view, in Fig. \ref{vn2}, the outage probability is provided for two different values of the UAV's instability ($\sigma_\theta=1^o$ and $1.4^o$) and also two different values of $\theta_{11}=20^o$ and $60^o$. 
As we expect, the system performance decreases by increasing $\sigma_\theta$. In addition, the simulation results show that by reducing $\theta_{11}$ from $60^o$ to $20^o$, the performance of the considered system improves. The reason for this is that by reducing $\theta_{11}$, the probability of having large $\theta_{d_{1j}}$ increases, and thus, the interference decreases and the system performance improves.

%

%

\section{Conclusion}
In this paper, we have studied mmWave downlink transmissions for an NFP-based network in the presence of real channel parameters. Our aim was to design a mmWave downlink and reuse frequency using directional antennas. In particular, we have performed a comprehensive modeling of inter-cell and intra-cell interference and by studying the effect of parameters, valuable solutions were provided to reduce the interference. From the obtained results, the upper bound of the outage probability was derived. Then, by providing numerical and simulation results, the effect of important channel parameters on the system performance was investigated and precious results were obtained, including the effect of integrating SBSs close to each other, how to choose the optimal antenna pattern to reduce interference and as well as reduce the negative effect of UAV fluctuations.

\appendices

\section{}  \label{AppA}
Based on the results of \cite{tripathi2021millimeter,dabiri2022pointing}, the main-lobe of the antenna pattern can be well approximated by Gaussian distribution. Therefore, \eqref{p6} can be rewritten as follows:
\begin{align}
	\label{c2}
	&P_{r_{11}} = P_t |h_{L_{11}}|^2 G_0(N_s) 
	G_0(N_u) \nonumber \\ 
	&\times \exp\left( -\frac{\left( \tan^{-1}\left(\sqrt{\tan^2(\Theta_{x_1})+\tan^2(\Theta_{y_1})}\right) \right)^2}{w_0^2}
	\right),
\end{align}
where $w_0=1/N_u$. 
Also, after a comprehensive search, we approximate the antenna pattern along with the side-lobes as a function of the sum of several Gaussian functions as:
\begin{align}
	\label{c5}
	G_u(N_u)         \simeq  G_0(N_u) \sum_{m=0}^M A_m A''_m \exp\left( -\frac{(\theta-A'_m)^2}{w_m^2}  \right)
\end{align}
where, $M=6$, and
\begin{align}
	\left\{ \!\!\!\!\! \! \!
	\begin{array}{rl}
		& w_m = \frac{1}{N_s}~~\&~A'_m=0,~~~~~~~~~~~~~~~~~~~~m=0 \\
		& w_m = \frac{1}{2N_s}~\&~A'_m=\sin^{-1}\left( \frac{2m+1}{N_u}\right),~~~~m\neq0 \\
	\end{array} \right. \nonumber
\end{align}
and $A_0=1, A_1=0.05, A_2=0.020, A_3=0.011, A_4=0.006, A_5=0.004$, $A_6=0.003$, and
$$\left\{ \!\!\!\!\! \! \!
\begin{array}{rl}
	& A''_m=1,~~\text{for}~~\frac{m}{N_u}\leq1 \\
	& A''_m=0,~~\text{for}~~\frac{m}{N_u}>1 \\
\end{array} \right. $$
In Fig. \ref{nv5}, for different values of $\phi=0:0.33:45$, the accuracy of the model obtained from \eqref{c5} is compared with the real antenna pattern model.
As we see, the approximated model provided in \eqref{c5} models well the main lobe along with the maximum amount of side-lobes.
We use \eqref{c5} to model the interference, and since the actual value of the interference is equal to or less than the interference obtained from \eqref{c5} (the actual value of the interference depends on RV $\phi_{u_i}$), the interference obtained from \eqref{c5} is for the worst scenario. Therefore, the outage probability obtained from \eqref{c5} will be the upper bound of the actual outage probability.
Now, from \eqref{p1}, \eqref{sd1}, \eqref{c2}, \eqref{c5} and using the results of { Remarks 3} and { 4}, the instantaneous SINR can be written as
\begin{align}
	\label{c6}
	& \gamma_{11} \simeq    \frac{\exp\left( -\frac{\left( \tan^{-1}\left(\sqrt{\tan^2(\Theta_{x_1})+\tan^2(\Theta_{y_1})}\right) \right)^2}{w_0^2}  \right)  }
	{\sum_{j=1}^J \sum_{m=0}^M A_m A''_m \exp\left( -\frac{(\theta_{d_{1j}}-A'_m)^2}{w_m^2}  \right) + N'_0 }
\end{align} 
where $N'_0= \frac{\sigma^2_{N}}{P_t |h_{L_{11}}|^2 G_0(N_s) G_0(N_u)}$ is the normalized thermal noise.
Let us define RV $\Theta'_1=\frac{\left( \tan^{-1}\left(\sqrt{\tan^2(\Theta_{x_1})+\tan^2(\Theta_{y_1})}\right) \right)^2}{w_0^2}$.
Note that $\Theta'_1$ is a random variable caused by the instability and vibrations of $U_1$, which is in the order of one degree (depending on the weather conditions, wind speed and the type of UAV, it can be a little more or less) \cite{dabiri20203d}. Therefore, since for small values of $x$ we have $\tan(x)\simeq x$, then $\Theta_1$ can be approximated as follows:
\begin{align}
	\label{c3}
	&\Theta'_1\simeq   \frac{\Theta_{x_1}^2+\Theta_{y_1}^2}{w_0^2}.
\end{align}
Using \cite[eqs. (3) and (5)]{nakagami1960m}, the distribution of $\Theta$ is obtained as
\begin{align}
	\label{c4}
	&f_{\Theta'_1}(\Theta_1) = \\& \frac{1}{\sqrt{\beta_{ux}\beta_{uy}}} \exp\left(-\frac{\beta_{ux}+\beta_{uy}}{2\beta_{ux}\beta_{uy}}\Theta'_1 \right)
	I_0\left( \frac{\beta_{ux}-\beta_{uy}}{2\beta_{ux}\beta_{uy}}\Theta'_1 \right) \nonumber
\end{align}
where $\beta_{uw}= \frac{\sqrt{2}\sigma_{\theta_w}^2}{w_0}$, and $I_0(\cdot)$ is the modified Bessel function of the first kind with order zero.
Using \eqref{c6} and \eqref{c4}, and after some derivations, the PDF of $\gamma_{11}$ is derived in \eqref{c7}.

\section{}  \label{AppB}
To calculate the outage probability, we need the CDF of $\Theta'_1$.
Based on \eqref{c7} and using \cite{reference.wolfram_2021_hoytdistribution}, the CDF of $\Theta'_1$ is derived as
\begin{align}
	\label{po8}
	F_{\Theta'_1}(\Theta'_1) &= Q\left( \mathbb{B}_1 \sqrt{\frac{2\Theta'_1}{\beta_{ux}+\beta_{uy}}}  ,  \mathbb{B}_2 \sqrt{\frac{2\Theta'_1}{\beta_{ux}+\beta_{uy}}}  \right)
	\nonumber \\ &~~~
	-  Q\left( \mathbb{B}_2 \sqrt{\frac{2\Theta'_1}{\beta_{ux}+\beta_{uy}}}  ,   \mathbb{B}_1 \sqrt{\frac{2\Theta'_1}{\beta_{ux}+\beta_{uy}}}  \right)
\end{align}
where $Q(a,b)$ is the Marcum {\it Q}-function, and 
\begin{align}
	\left\{ \!\!\!\!\! \! \!
	\begin{array}{rl}
		&\mathbb{B}_1 = \frac{\sqrt{1-T_q^4}}{2T_q} \sqrt{\frac{1+T_q}{1-T_q}},\\
		&\mathbb{B}_2 = \mathbb{B}_1 \frac{1-T_q}{1+T_q},   ~~~ 
		T_q  = \frac{\sigma_{\theta_\text{min}}}{\sigma_{\theta_\text{max}}}, \\
		&\sigma_{\theta_\text{max}} = \text{max}\{\sigma_{\theta_{x}},\sigma_{\theta_{y}}\},~~~~
		\sigma_{\theta_\text{min}} = \text{min}\{\sigma_{\theta_{x}},\sigma_{\theta_{y}}\}
	\end{array} \right. \nonumber
\end{align}
Using \eqref{c6} and \eqref{po8}, and after some derivations, the CDF of $\gamma_{11}$ is derived in \eqref{c8}.

\section{}  \label{AppC}
Here, we want to derive the distribution of $\theta_{11}=[\theta_{x_{11}},\theta_{y_{11}}]$. Using \eqref{f_2}, it can be formulated as
\begin{align}
	\label{pk4}
	\theta_{11}  &= \tan^{-1}\left(\sqrt{\tan^2(\theta_{x_{11}})+\tan^2(\theta_{y_{11}})}\right).
\end{align}
Considering that each UAV can cover SBSs up to a maximum distance of $L_\text{max}$, so for $U_1$ placed at $[0,0,h_1]$ in a Cartesian coordinate system $[x,y,z]\in\mathbb{R}^{1\times3}$, the SBSs located in a circular plane with a radius of $R$ can be connected to $U_1$ where
$R=\sqrt{L_\text{max}^2-h_1^2}$.
As mentioned earlier, the position of $\mathcal{S}_{1j}$ is characterized by $[x_{1j},y_{1j},0]$ in the considered Cartesian coordinate system where $x_{1j}$ and $y_{1j}$ are uniform RV that 
\begin{align}
	\label{pk1}
	f_{x_{1j}}(x_{1j}) = \frac{1}{2R},~~~\text{for}~~~-R<x_{1j}<R,
\end{align}
and
\begin{align}
	\label{pk2}
	f_{y_{1j}|x_{1j}}(y_{1j}) = \frac{1}{2Y_\text{max}}~~~\text{for}~~~-Y_\text{max}<y_{1j}<Y_\text{max}.
\end{align}
In \eqref{pk2}, we have $Y_\text{max}=\sqrt{R^2-x_{1j}^2}$.
Based on \eqref{p5}, we have
\begin{align}
	\label{pk3}
	\theta_{x_{1j}}=\tan^{-1}\left(\frac{x_{1j}}{h_1}\right),~~~~\&~~~~\theta_{y_{1j}}=\tan^{-1}\left(\frac{y_{1j}}{h_1}\right).
\end{align}
Using \eqref{pk1} and \eqref{pk3}, the PDF of $\theta_{x_{1j}}$ is obtained as
\begin{align}
	\label{pk5}
	F_{\theta_{x_{11}}}(\theta_{x_{11}}) &= \text{Prob}\left\{ \tan^{-1}\left( \frac{x_{11}}{h_1} \right)< \theta_{x_{11}} \right\} \nonumber \\
	&=F_{x_{11}}\left( h_1 \tan\left(\theta_{x_{11}}\right)  \right).
\end{align}
Without loss of generality, we assume that $\mathcal{S}_{11}$ is placed on the $x$-axis. Therefore, based on \eqref{pk4}, we have $\theta_{11}=|\theta_{x_{11}}|$ and thus:
\begin{align}
	\label{pk6}
	f_{\theta_{{11}}} (\theta_{11}=|\theta_{x_{11}}|) = \frac{\text{d}F_{|\theta_{x_{11}}|}(|\theta_{x_{11}}|)}{\text{d}|\theta_{x_{11}}| },
	~~~ 
    0\! <\!\theta_{11}\!<\!\tan^{-1}\left(\!\frac{R}{h_1}\!\right) 
\end{align}
Finally, using \eqref{pk5} and \eqref{pk6}, the PDF of $\theta_{11}$ is derived in \eqref{pk7}.

%

\section{}  \label{AppD}
In this appendix, we derive the distribution of $\theta_{d_{1j}}$. Based on the assumption of Appendix \ref{AppC}, $\theta_{d_{1j}}$ can be formulated as
\begin{align}
	\label{pk8}
	\theta_{d_{1j}} = \tan^{-1}\left(\sqrt{\tan^2(\theta_{x_{1j}}-\theta_{x_{11}})+\tan^2(\theta_{y_{1j}})}\right).
\end{align}
Based on \eqref{pk8}, the CDF of $\theta_{d_{1j}}$ conditioned on $\theta_{x_{1j}}$ is obtained as
\begin{align}
	\label{pk9}
	&F_{\theta_{d_{1j}} | \theta_{x_{1j}}}  \\
	&\resizebox{0.99\hsize}{!}{$
	= \text{Prob}\left\{ \tan^{-1}\left(\sqrt{\tan^2(\theta_{x_{1j}}-\theta_{x_{11}})+\tan^2(\theta_{y_{1j}})}\right)
	< \theta_{d_{1j}} \bigg| \theta_{x_{1j}}  \right\}        $}\nonumber \\
	&\resizebox{0.99\hsize}{!}{$
	=\text{Prob}\left\{ \theta_{y_{1j}}
	< \tan^{-1}\left( \sqrt{\tan^2(\theta_{d_{1j}})  -  \tan^2(\theta_{x_{1j}}-\theta_{x_{11}}) }  \right) \bigg| \theta_{x_{1j}}  \right\}   
	$}\nonumber \\
    &\!\!=F_{\theta_{y_{1j}}  | \theta_{x_{1j}} }\left(
    \tan^{-1}\left( \sqrt{\tan^2(\theta_{d_{1j}})  -  \tan^2(\theta_{x_{1j}}-\theta_{x_{11}}) }  \right)
    \right). \nonumber
\end{align}
The CDF of $\theta_{y_{1j}}$ is obtained as 
\begin{align}
	\label{nk5}
	F_{\theta_{y_{1j}}}(\theta_{y_{1j}}) = \int F_{\theta_{y_{1j}}  | \theta_{x_{1j}} }(\theta_{y_{1j}}) 
	f_{ \theta_{x_{1j}} }(\theta_{x_{1j}}) \text{d} \theta_{x_{1j}}.
\end{align} 
Following the results of Appendix \ref{AppC}, the PDF of $\theta_{x_{1j}}$ and the CDF of $\theta_{y_{1j}}$ conditioned on $\theta_{x_{1j}}$ are derived  respectively as
\begin{align}
	\label{nk4}
	\resizebox{0.99\hsize}{!}{$
	f_{ \theta_{x_{1j}} }(\theta_{x_{1j}}) = 
	\frac{h_1}{R \cos^2(\theta_{x_{1j}})},~~~
	-\tan^{-1}\left(\frac{R}{h_1}\right)<\theta_{x_{1j}}<\tan^{-1}\left(\frac{R}{h_1}\right),
	$}
\end{align}
\begin{align}
	\label{nk1}
	F_{\theta_{y_{1j}}  | \theta_{x_{1j}} }(\theta_{y_{1j}}) = \frac{h_1}{\sqrt{R^2-h_1^2\tan^2(\theta_{x_{1j}})}}  \tan(\theta_{y_{1j}}).
\end{align}
From \eqref{pk9} and \eqref{nk1}, we obtain
\begin{align}
	\label{nk2}
	&F_{\theta_{d_{1j}} | \theta_{x_{1j}}} (\theta_{d_{1j}}) = \\
	&\frac{h_1}{\sqrt{R^2-h_1^2\tan^2(\theta_{x_{1j}})}}
	\sqrt{\tan^2(\theta_{d_{1j}})  -  \tan^2(\theta_{x_{1j}}-\theta_{x_{11}}) }. \nonumber
\end{align}
By derivation from \eqref{nk2}, the PDF of $\theta_{d_{1j}}$ conditioned on $\theta_{x_{1j}}$ is derived in \eqref{nk3}.

\section{}  \label{AppE}
We have previously obtained outage probability condition on $\theta_{d_{1j}}$ in Theorem 2. Using \eqref{c8}, the average outage probability of the considered system is obtained as:
\begin{align}
	\label{vm1}
	\mathbb{P}_{\text{out}} = \int ...\int \mathbb{P}_{\text{out}|\theta_{d_{1j}}} f_{\theta_{d_{1j}}}(\theta_{d_{1j}}) \text{d}\theta_{d_{12}}...\text{d}\theta_{d_{1J}}.
\end{align}
Outage probability of the considered system is computed by substituting \eqref{c8} and \eqref{nk6} in \eqref{vm1}, which requires to solve a $2(J-1)$ dimensional numerical integration.
Since \eqref{c8} and \eqref{nk6} are approximately complex, solving the above $2(J-1)$ dimensional integral will be very complicated if not impossible.
In the following, we use the results of Theorem 2 as well as Propositions 1 and 2, and by removing ineffective terms, we simplify \eqref{vm1}.

Based on the results of Theorem 2, by increasing $\theta_{d_{1j}}$, the average outage probability decreases. In other words, to calculate the average outage probability, we can only consider the effect of $\mathcal{S}_{1j}$ close to $\mathcal{S}_{11}$ that have the smallest $\theta_{d_{1j}}$.
For better understanding, a numerical example is provided. For example, suppose that for 4 random $\mathcal{S}_{1j}$, the random value of spatial angle $\theta_{d_{1j}}$ are equal to $20^o$, $40^o$, $100^o$, and $140^o$, which respectively causes average outage probability of $10^{-3}$, $10^{-5}$, $10^{-7}$, and $10^{-8}$, respectively. It can be easily seen that the final outage probability is dominated to the closest $\mathcal{S}_{1j}$, because we have $(10^{-3}+10^{-5}+10^{-7}+10^{-8})\simeq10^{-3}$.
To find the SBSs with the smallest $\theta_{d_{1j}}$, we sectorize the parameter $\theta_{d_{1j}}$ as:
\begin{align}
	\theta_{d_{1j}}\simeq \frac{d\mathbb{G}_2}{D}+\mathbb{G}_1, ~~~\text{for}~~~ \frac{d\mathbb{G}_2}{D}<\theta_{d_{1j}}-\mathbb{G}_1<\frac{(d+1)\mathbb{G}_2}{D},
	\nonumber
\end{align}
where $d\in\{0,1,...,D\}$, $\mathbb{G}_1=\max\{\alpha_c,|\theta_{x_{1j}}-\theta_{x_{11}}|\}$, and $\mathbb{G}_2 = (\theta_{d\text{max}}-\mathbb{G}_1)$.
For obtaining outage probability, we consider the effect of the $\mathcal{S}_{1j}$ located in the nearest angular sectors. 
It is clear that as $D$ increases, the accuracy of the derivations increases.
In the simulations, we show that $D=80$ provides acceptable accuracy. According to the mentioned points, the $2(J-1)$-dimensional integration \eqref{vm1} is simplified to the following one-dimensional integral:
\begin{align}
	\label{vm2}
	\mathbb{P}_{\text{out}} = \int \mathbb{P}_{\text{out}|\theta_{x_{1j}}} 
	f_{\theta_{x_{1j}}} (\theta_{x_{1j}})              \text{d}\theta_{x_{1j}}.
\end{align}
where 
\vspace{-.5 cm}
\begin{align}
	\label{vm3}
	&\mathbb{P}_{\text{out}|\theta_{x_{1j}}} = \sum_{d=0}^{D-1}\sum_{j=1}^{J-1}\binom{J}{j}
	\mathbb{P}_{\text{out}|\frac{d\mathbb{G}_2}{D}} \\
	& \times
	\left( \text{Prob}\left\{\frac{d}{D}<\frac{\theta_{d_{1j}}-\mathbb{G}_1}{\mathbb{G}_2}<\frac{d+1}{D}
	\Big| \theta_{x_{1j}} \right\} \right)^j \\
	& \times
	\left( \text{Prob}\left\{\frac{d+1}{D}<\frac{\theta_{d_{1j}}-\mathbb{G}_1}{\mathbb{G}_2}   \Big| \theta_{x_{1j}}   \right\} \right)^{J-j-1}.
\end{align}
In \eqref{vm3}, $\binom{n}{m}$ is the combinations of $m$ and $n$.
Also, for each $d$, the term $\left( \text{Prob}\left\{\frac{d}{D}<\frac{\theta_{d_{1j}}-\mathbb{G}_1}{\mathbb{G}_2}<\frac{d+1}{D}
\Big| \theta_{x_{1j}} \right\} \right)^j $ is proportional to the probability of placing $j$ of $J-1$ SBSs in the $d$th sector, and the term $ \left( \text{Prob}\left\{\frac{d+1}{D}<\frac{\theta_{d_{1j}}-\mathbb{G}_1}{\mathbb{G}_2}   \Big| \theta_{x_{1j}}   \right\} \right)^{J-j-1}$ is proportional to the probability of placing the rest of the SBSs in sectors with $\theta_{d_{1j}}>\left(\frac{(d+1)\mathbb{G}_2}{D}+\mathbb{G}_1\right)$.
Finally, using \eqref{nk8}, \eqref{nk9}, \eqref{c8}, and \eqref{vm3}, and after a series of derivations, the average outage probability of the considered system is derived in \eqref{bn1} and \eqref{pn2}.


\end{document}